\documentclass[12pt]{article} 
\usepackage{amssymb}
\usepackage{amsmath} 
\usepackage{cite}
\usepackage{hyperref}
\input xy
\xyoption{all}

\numberwithin{equation}{section}

\newcommand{\bC}{\mathbb{C}}
\newcommand{\dd}{\mathrm{d}}
\newcommand{\fh}{\mathfrak{h}}
\newcommand{\fM}{\mathfrak{M}}
\newcommand{\sQ}{\mathsf{Q}}

\DeclareMathOperator*{\JKres}{JK-Res}
\DeclareMathOperator{\rank}{rank}
\newcommand{\Tr}{{\rm Tr}}

\begin{document} 

\begin{flushright}
KIAS-P20047\\
\end{flushright}
\begin{center}

{\large\bf Elliptic genera of pure gauge theories in two dimensions\\
with semisimple non-simply-connected gauge groups}

\vspace{0.2in}

Richard Eager$^1$, Eric Sharpe$^2$

\begin{tabular}{cc}
{\begin{tabular}{l}
$^1$ School of Physics\\
Korea Institute for Advanced Study\\
Seoul 02455, Korea
\end{tabular}} &
{\begin{tabular}{l}
$^2$ Dep't of Physics\\
Virginia Tech\\
850 West Campus Dr.\\
Blacksburg, VA  24061
\end{tabular}}
\end{tabular}

{\tt reager@kias.re.kr}, {\tt ersharpe@vt.edu}

Abstract
\end{center}

In this paper we describe a systematic method
to compute elliptic genera of 
(2,2) supersymmetric 
gauge theories
in two dimensions
with gauge group
$G/\Gamma$ (for $G$ semisimple and simply-connected,
$\Gamma$ a subgroup of the center of $G$) with various discrete
theta angles.  We apply the technique to examples of pure gauge theories
with low-rank gauge groups.  Our results are consistent with expectations from
decomposition of two-dimensional theories with finite global one-form
symmetries and with computations of supersymmetry breaking for some
discrete theta angles in
pure gauge theories.
Finally, we make predictions for the elliptic genera of all the other remaining pure gauge theories
by applying decomposition and matching to known supersymmetry breaking patterns.

\begin{flushleft}
September 2020
\end{flushleft}

\newpage

\tableofcontents

\newpage

\section{Introduction}

The low energy infrared (IR) limits of gauge theories have been of interest for many years.
Pure gauge theories in two dimensions with $\mathcal{N} = (2,2)$ supersymmetry
have long been believed to be gapless, 
as a result of the chiral R-symmetry and anomalous
two-point functions \cite[section 3]{Witten:1995im}.
The paper \cite{Aharony:2016jki} made a more refined conjecture: that the IR
limit of a (2,2) supersymmetric pure $G$ gauge theory, $G$ semisimple
and simply-connected,
should be a theory
of free twisted chiral multiplets, as many as the rank of $G$, with
R-charges
proportional to
Casimir degrees.  Using nonabelian mirrors
\cite{Gu:2018fpm} it was checked in \cite{Gu:2018fpm,Chen:2018wep,Gu:2020ivl} 
that the IR theory contains as many twisted chirals as the rank,
and in pure $G/\Gamma$ gauge theories for $\Gamma$ a subgroup of the center of $G$,
that one gets an identical free theory for one value of the discrete theta
angle, and supersymmetry breaking in the IR for other values of the discrete
theta angle.

All that said, the work \cite{Gu:2018fpm,Chen:2018wep,Gu:2020ivl} did not
compute elliptic genera, which would provide a very explicit 
concrete check of R-charges of free IR twisted chirals.  For a pure
(2,2) supersymmetric $G$ gauge theory for $G$ simply-connected,
methods to compute elliptic genera exist (see e.g. 
\cite{Benini:2013nda,Benini:2013xpa,Gadde:2013dda,Kim:2014dza}),
and it is being checked \cite{Eager:2019bxq, Eager}, that those elliptic genera match the
expectations of \cite{Aharony:2016jki}.

The purpose of this
paper is to develop the technology to compute elliptic genera of pure
(2,2) supersymmetric $G/\Gamma$ gauge theories for various discrete theta angles.
The elliptic genus is given by a sum of Jeffrey--Kirwan residues of a meromorphic form over the moduli space of
flat $G/\Gamma$-connections the torus using supersymmetric localization \cite{Benini:2013nda,Benini:2013xpa}.
The meromorphic form is obtained by evaluating the one-loop determinants corresponding to
$G/\Gamma$-bundles with non-trivial characteristic classes. 
We combine the results from different components of the moduli space, weighted by phases from the discrete theta angle, to determine the elliptic genus.

In section~\ref{sect:rev} we review known results for
elliptic genera of pure supersymmetric gauge theories in two dimensions.
In section~\ref{sect:strategy} we describe the procedure we will use
to compute elliptic genera of pure supersymmetric gauge theories with
semisimple but non-simply-connected gauge groups.
The remainder of this paper is spent working out low-rank examples.
We begin in section~\ref{sect:so3} by discussing pure $SO(3)$ gauge theories.
For these, the elliptic genera in question were previously derived
in \cite[appendix A]{Kim:2017zis}, 
but this case acts as a test and demonstration of
our strategy.
In section~\ref{sect:su3}, we compute elliptic genera of
pure $SU(3)/{\mathbb Z}_3$ gauge theories;
in section~\ref{sect:so4}, pure $SO(4)$ gauge theories;
in section~\ref{sect:spin4}, pure Spin$(4)/({\mathbb Z}_2 \times
{\mathbb Z}_2)$ gauge theories;
in section~\ref{sect:so5}, pure $SO(5)$ gauge theories;
and, in section~\ref{sect:sp6}, pure
$Sp(6)/{\mathbb Z}_2$ gauge theories.
In each case, 
the
elliptic genus vanishes (and supersymmetry is broken)
unless the discrete theta angle takes the
value described in \cite{Gu:2018fpm,Chen:2018wep,Gu:2020ivl}.
We conclude by making predictions for elliptic genera of all other
pure gauge theories with semisimple non-simply-connected gauge groups,
in section~\ref{sect:predict}.

We will also note in each case that the results are consistent with
decomposition \cite{Hellerman:2006zs,Sharpe:2014tca,Sharpe:2019ddn}.
(See also e.g. \cite{Tanizaki:2019rbk,Komargodski:2020mxz} for
four-dimensional versions and related analyses.)  Briefly, decomposition
is the statement that a two-dimensional theory with a finite global
1-form symmetry (such as a two-dimensional gauge theory in which 
a finite center acts trivially) decomposes\footnote{
This is a stronger statement than just superselection.
For example, only in infinite volume does one get a selection
rule from superselection sectors, whereas decomposition holds at finite volume.
This distinction is discussed in
greater detail in \cite{Tanizaki:2019rbk}.
} into a disjoint union of
theories which individually do not have a 1-form symmetry.  In the
case of a pure $G$ gauge theory for $G$ simply-connected, with $\Gamma$ a finite
subgroup of the center, the $G$ gauge theory has a global 
one-form $\Gamma$ symmetry (sometimes denoted $B\Gamma$),
and so decomposes into a disjoint union of $G/\Gamma$ gauge theories with
various discrete theta angles, which we write schematically as
\begin{equation}
G \: = \: \oplus_{\theta \in \hat{\Gamma}} \left( G/\Gamma \right)_{\theta}.
\end{equation}
In particular, the elliptic genus of a pure $G$ gauge theory should
be the sum of elliptic genera of pure $G/\Gamma$ gauge theories with various
discrete theta angles.  The result that the elliptic genera of 
$G/\Gamma$ gauge theories vanish except for a single discrete theta angle,
for which the elliptic genus matches that of the $G$ gauge theory,
is consistent with the decomposition above.

Also, although we will not emphasize this perspective in this paper,
in principle these computations have a mathematical understanding.
Elliptic genera of pure $G$ gauge theories should, in principle,
match 
\cite{Pantev:2005rh,Pantev:2005zs,Pantev:2005wj}
elliptic genera of classifying stacks $BG$,
the $G$-equivariant elliptic genera of points \cite{gro,bet},
and so we are also making predictions for those elliptic genera.

\section{Review and overview}
\label{sect:rev}

Pure $\mathcal{N} = (2,2)$ supersymmetric $G$ gauge theory can be described in terms of
vector multiplet consisting of a gauge field $A_\mu,$ gauginos $\lambda$ and $\bar\lambda$, scalars $\sigma, \bar\sigma,$ and a real auxiliary scalar $D.$
The gauge field strength is a twisted chiral superfield $\Sigma$ with lowest component $\sigma.$
The Euclidean Yang-Mills Lagrangian is
\begin{equation}
\mathcal{L}_\text{YM} = \Tr \Big( F_{12}^2 + D^2 + D_\mu \bar\sigma D^\mu \sigma + i D [\sigma, \bar\sigma] - i \bar\lambda \gamma^\mu D_\mu \lambda - i \bar\lambda P_+ [\sigma, \lambda] - i \bar\lambda P_- [\bar\sigma, \lambda] \Big) \;,
\end{equation}
where 
\begin{equation}
P_\pm = \frac{1 \pm \gamma_3}{2}.
\end{equation} 
The classical potential is proportional to $\Tr \left[ \sigma, \sigma^{\dagger} \right]^2.$  The classical vacua occur at the minimum of the potential and satisfy
$\left[ \sigma, \sigma^{\dagger} \right] = 0.$
Equivalently, the classical Coulomb branch of vacua can be described by the vacuum expectation values of the gauge invariant polynomials in $\sigma$.  It is a classical result that this ring of
functions is freely generated by $\text{rank}(G)$ generators.
However, the potential receives quantum corrections, so the IR behavior
is potentially more complex.

\subsection{Prediction for simply-connected semisimple $G$}

The paper \cite{Aharony:2016jki} proposed that for $G$ semisimple
and simply-connected, the
IR theory should be a free theory of twisted chiral multiplets, $Y_i(\Sigma)$, $i = 1, \dots, \text{rank}(G),$
built out of the generators of the invariant functions on $\Sigma$, with
axial R-charges $r_i$ given by
twice the Casimir degrees\footnote{
This follows from the Harish-Chandra isomorphism that relates Casimirs to 
symmetric invariants.
} $d_i$ of $G$
computed from and in one-to-one correspondence with the possible
Casimirs (of which there are as many as the rank).
The contribution of a single twisted chiral multiplet $Y(\Sigma)$  with 
axial R-charge $r$ to the elliptic genus is \cite[equ'n (2.11)]{Benini:2013xpa}
\begin{equation} 
\label{twisted chiral}
{\rm Tr}_{RR} \,  (-1)^F q^{H_L} \overline{q}^{H_R} y^J \: = \: 
\frac{ \theta_1( \tau | (1- r/2) z ) }{ \theta_1( \tau |  - (r/2) z) },
\end{equation}
where $q = \exp(2 \pi i \tau)$, $y = \exp(2 \pi i z)$, $J$ is the left-moving $U(1)_R$ charge,
and the genus is computed for periodic left-moving fermions.
Since the low energy theory is a theory of free twisted chiral multiplets, the elliptic genus is expected to be
\begin{equation}  \label{eq:ellgen-pred}
\prod_i \frac{ \theta_1( \tau | (1- r_i/2) z ) }{ \theta_1( \tau |  - (r_i/2) z) }.
\end{equation}
For simply-connected $G$, this will be demonstrated by explicit
computation in \cite{Eager}.

For later use, we collect in table~\ref{table:casimirs}
the degrees of Casimirs for simple Lie
algebras, each of which is half the R-charge of a corresponding
twisted chiral in equation~(\ref{eq:ellgen-pred}).  
For example, the elliptic genus of a pure $G_2$ gauge theory is
predicted to be
\begin{equation}
\frac{ \theta_1( \tau | - z) }{ \theta_1( \tau | - 2z) }
\frac{ \theta_1( \tau | -5 z) }{ \theta_1( \tau | -6 z) }.
\end{equation}
As a consistency check, the
Casimir degrees $d_i$ and the dimension of the group $G$ are related by
\begin{equation}
{\rm dim}\, G \: = \: \sum_i (2 d_i - 1).
\end{equation}

In passing, identifying R-charges $r_i = 2 d_i$,
we can apply the central charge formula\footnote{
In conventions in which the superpotential obeys
$W(\lambda^{r_i} x_i) \: = \: \lambda^2 W(x_i)$.
} \cite{Vafa:1989xc}[equ'n (15)]
to see that
\begin{eqnarray}
\frac{c_{\rm eff}}{3} & = & \sum_i \left( 1 - r_i \right)
\: = \: - {\rm dim}\, G,
\end{eqnarray}
where $c_{\rm eff}$ is an effective central charge,
differing from the ordinary central charge as 
\cite[equ'n (13)]{Itzykson:1986pk}
\begin{equation}
c_{\rm eff} \: = \: c - 24 h_{\rm min},
\end{equation}
for $h_{\rm min}$ the smallest conformal dimension appearing in the theory,
as relevant to theories with continuous spectra
\cite{Haghighat:2015ega,Cecotti:2015lab}.
We can get the same result from the modular transformation properties.
Applying \cite[equ'n (2.7)]{Benini:2013xpa}
\begin{equation}
Z\left( - \frac{1}{\tau} , \frac{z}{\tau} \right)
\: = \: 
\exp\left[ \frac{c_{\rm eff}}{3} \frac{\pi i}{\tau} z^2 \right]
Z( \tau, z)
\end{equation}
and the modular transformation property
\cite[equ'n (A.8)]{Benini:2013xpa}
\begin{equation}
\theta_1\left( \left. - \frac{1}{\tau} \right| \frac{z}{\tau} \right)
\: = \: -i \sqrt{-i \tau} \exp(\pi i z^2/\tau) \theta_1(\tau | z),
\end{equation}
we see that under $\tau \mapsto -1/\tau$, $z \mapsto z/\tau$,
the elliptic genus of a twisted chiral with R-charge $r$
(equ'n~(\ref{twisted chiral})) picks up a phase
\begin{equation}
\exp\left( \pi i (1-r) z^2/\tau \right),
\end{equation}
and the elliptic genus of a pure $G$ gauge theory~(\ref{eq:ellgen-pred})
picks up a phase
\begin{equation}
\exp\left( \pi i \sum_i (1-r_i) z^2/\tau \right) \: = \:
\exp\left( - \pi i ({\rm dim}\, G) z^2/\tau \right).
\end{equation}
This phase is determined by the smallest conformal weight $h_{\rm min}$
appearing
in the theory, following \cite{Itzykson:1986pk}.
In any event,
we see again that the (left-moving) effective central charge is given by
\begin{equation}
\frac{c_{\rm eff}}{3} \: = \: - {\rm dim}\, G.
\end{equation}
Intuitively, for theories formulated on $S^2$, the sign of the
central charge above is surely related
to the fact that for R charge greater than two, the action has a 
curvature-dependent term of the wrong sign
\cite[section 3.4]{Hori:2013ika}.

Mathematically, this has a simple understanding.  A pure $G$-gauge
theory is a sigma model on \cite{Pantev:2005rh,Pantev:2005zs,Pantev:2005wj}
the stack $BG = [{\rm point}/G]$,
and this stack has dimension (see e.g. \cite[section 7]{vistoli},
\cite[example 2.44]{gomez})
\begin{equation}
{\rm dim}\, [ {\rm point}/G] \: = \: - {\rm dim}\, G,
\end{equation}
matching $c_{\rm eff}/3$ above.

\begin{table}
\begin{center}
\begin{tabular}{ccc}
Gauge group & Dimension & Casimir degrees \\ \hline
$SU(n+1) (A_n)$ & $(n+1)^2-1$ & $2, 3, 4, \cdots, n+1$ \\
Spin$(2n+1) (B_n)$ & $n(2n+1)$  & $2, 4, 6, \cdots, 2n$ \\
$Sp(2n) (C_n)$ & $n(2n+1)$ & $2, 4, 6, \cdots, 2n$ \\
Spin$(2n) (D_n)$ & $n(2n-1)$ & $n; 2, 4, 6, \cdots, 2n-2$ \\
$G_2$ & $14$ & $2, 6$ \\
$F_4$ & $52$ & $2, 6, 8, 12$ \\
$E_6$ & $78$ & $2, 5, 6, 8, 9, 12$ \\
$E_7$ & $133$ & $2, 6, 8, 10, 12, 14, 18$ \\
$E_8$ & $248$ & $2, 8, 12, 14, 18, 20, 24, 30$
\end{tabular}
\end{center}
\caption{List of Casimir degrees for various gauge groups, each corresponding
to half an R-charge.
See e.g. \cite[table 5a]{schwarz78}.
\label{table:casimirs} }
\end{table}

\subsection{Non-simply-connected $G$}
\label{sect:rev:nsc}

In this paper, we will compute elliptic genera of pure supersymmetric
gauge theories with gauge groups $G/\Gamma$, where $G$ is simply-connected
and $\Gamma$ is a subgroup of the center of $G$.  Now, a principal $G/\Gamma$
bundle on worldsheet $T^2$ admits a characteristic class we shall
denote $w \in H^2(T^2,\Gamma) \cong \Gamma$.  (For example, for $SO(k)$ bundles,
$w$ is the Stiefel-Whitney class $w_2$.)  Such theories admit
analogues of theta angles, known as discrete theta angles,
in which the path
integral is weighted by phases of the form $\exp(i \theta \cdot w)$ for
$\theta$ a (log of a) character of $\Gamma$, the set of which we shall denote
$\hat{\Gamma}$.

The papers \cite{Gu:2018fpm,Chen:2018wep,Gu:2020ivl}
have looked at IR behavior of two-dimensional pure (2,2) supersymmetric
gauge theories with non-simply-connected gauge
groups $G/\Gamma$.  (See also \cite{Kim:2017zis,Avraham:2019uiz,Bergman:2018vqe} 
for computations of elliptic genera
in some examples related to Hori's dualities \cite{Hori:2011pd}.)
Briefly, these papers found
\begin{itemize}
\item If the gauge group is not simply-connected, then for precisely one value
of the discrete theta angle, the IR limit is a theory of free twisted
chirals, as many as the rank (and as many as IR limit of a pure
gauge theory with corresponding simply-connected gauge group).
For other values of the discrete theta angle, there are no supersymmetric vacua,
hence supersymmetry is broken in the IR.
\item For the one nontrivial case, the IR theory is a theory of as many
twisted chiral multiplets as the rank, matching the IR behavior
of a pure $G$ gauge theory.
\end{itemize}
This structure is consistent with the predictions of decomposition
\cite{Hellerman:2006zs,Sharpe:2014tca,Sharpe:2019ddn}
for two-dimensional theories with one-form symmetries,
as discussed in \cite{Gu:2018fpm,Chen:2018wep,Gu:2020ivl}.

In this paper, we will compute elliptic genera to check these claims for
more general theories.

\begin{table}
\begin{center}
\begin{tabular}{cc}
Gauge group & Discrete theta angle for which susy unbroken \\ \hline
$SU(k)/{\mathbb Z}_k$ & $- (1/2) k (k-1) \mod k$ \\
Spin$(2k+1)/{\mathbb Z}_2$ & $1 \mod 2$ \\
Spin$(4k)/{\mathbb Z}_2 \times {\mathbb Z}_2$ &
$k (2k-1) \mod 2,$ $0 \mod 2$ \\
Spin$(4k+2)/{\mathbb Z}_4$ & $2k(2k-1) \mod 4$ \\
$Sp(2k)/{\mathbb Z}_2$ & $(1/2) k (k+1) \mod 2$ \\
$E_6/{\mathbb Z}_3$ & $0 \mod 3$ \\
$E_7/{\mathbb Z}_2$ & $1 \mod 2$
\end{tabular}
\end{center}
\caption{List of distinguished discrete theta angles for various 
non-simply-connected gauge groups,
for which a pure gauge theory admits supersymmetric vacua, summarizing
results from \cite{Gu:2018fpm,Chen:2018wep,Gu:2020ivl}.
\label{table:dta} }
\end{table}

To understand some of the quantum subtleties that will arise when studying
pure $G/\Gamma$ gauge theories, let us briefly review such theories
more concretely.  The Lagrangian for such
a theory can be written in (2,2) superspace in the form\footnote{
See e.g. \cite[section 4.1]{Witten:1993xi}.
}
\begin{equation} \label{eq:twistedaction}
- \frac{1}{4 g^2} \int d^4 \theta {\rm Tr}\, 
\overline{\Sigma} \Sigma \: + \: 
\left( - r + i \frac{\theta}{2\pi} \right)
\int d \theta^+ d \overline{\theta}^-
\left. {\rm Tr} \Sigma \right|_{\theta^- = \overline{\theta}^+ = 0}
\: + \: c.c.,
\end{equation}
where $\Sigma$ is a twisted chiral superfield encoding the gauge
field strength, $r$ is a Fayet-Iliopoulos parameter,
and $\theta$ the theta angle.
In analyzing the low-energy behavior of such
theories one
often works on the Coulomb branch, along which there is a
twisted one-loop effective superpotential which for a pure
$G/\Gamma$ gauge theory with $G$ simply-connected and $\Gamma$ a subgroup of
the center, takes the form
\begin{equation}
W_{\rm eff} \: = \: - \sum_a \Sigma_a \left[ - r_a + i \frac{\theta_a}{2\pi} 
\: + \: \frac{1}{|\Gamma|}
\sum_{\tilde{\mu}} \alpha^a_{\tilde{\mu}} \left( 
\ln \left( \sum_b \Sigma_b \alpha^b_{\tilde{\mu}} \right) - 1 
\right) \right],
\end{equation}
where now $r_a$ and $\theta_a$ are the FI parameters and theta
angles for each of the unbroken $U(1)$'s on the Coulomb branch.
(No further corrections exist beyond one-loop order.)
The first two terms are the $(-r + i \theta/2\pi) {\rm Tr} \Sigma$ 
of the classical
action along the Coulomb branch,
and the last is a loop correction, of the same form
commonly seen in theories with matter, here ultimately due to W bosons.
The $\alpha^a_{\tilde{\mu}}$ are the root vectors of the nonzero roots
(indexed by $\tilde{\mu}$)
of the Lie algebra of the gauge group.  The second term can be
simplified, and written as (see e.g. \cite[section 2.1]{Gu:2020ivl})
\begin{equation}
\frac{1}{|\Gamma|}
\sum_{\tilde{\mu}} \alpha^a_{\tilde{\mu}} \left(
\ln \left( \sum_b \Sigma_b \alpha^b_{\tilde{\mu}} \right)
- 1 \right)
\: = \:
\sum_{\tilde{\mu} \: {\rm pos'}} \frac{i \pi}{|\Gamma|} \alpha^a_{\tilde{\mu}},
\end{equation}
giving what amounts to a gauge-group-dependent shift of the theta angle.
(This was first observed in \cite[equ'n (10.9)]{Hori:2013ika}.)
These additional phases will play an important role in our
computations of elliptic genera of pure $G/\Gamma$ gauge theories.

\section{Strategy to compute elliptic genera}
\label{sect:strategy}

The elliptic genus of a pure $G/\Gamma$-gauge theory reduces to a residue integral 
over the moduli space ${\cal M}$ of flat $G/\Gamma$-connections on $T^2$ \cite{Benini:2013nda, Benini:2013xpa}.
Principal $G/\Gamma$ bundles have a degree-two characteristic
class, valued in $\Gamma$, which we shall denote $w \in H^2(T^2,\Gamma) \cong \Gamma$,
so the moduli space of flat $G/\Gamma$ connections is a disjoint union of moduli spaces
\begin{equation}
{\cal M} = \bigsqcup_{w \in H^2(T^2,\Gamma)} {\cal M}_{G/\Gamma,w}. 
\end{equation}

In the sector of bundles with $w=0$, any $G/\Gamma$ bundle lifts to a $G$
bundle.  Essentially as a result, the elliptic genus of a pure $G$
gauge theory matches that of a pure $G/\Gamma$ gauge theory in the sector
$w=0$, up to a volume factor $1/|\Gamma \times \Gamma|$ and a Jacobian factor $|\Gamma|$\footnote{This arises from the different normalization of the root systems.}:
\begin{equation}
\label{eq:factor}
Z(G/\Gamma, w=0) \: = \: \frac{|\Gamma|}{|\Gamma \times \Gamma|} Z(G) = \frac{1}{|\Gamma|} Z(G) .
\end{equation}

Now, we turn to a $G/\Gamma$ gauge theory in a sector in which $w \neq 0$.
Computations in these sectors will occupy most of the effort in this paper.
To describe such bundles, we pick two holonomies $p$, $q$ around
cycles of the torus, which commute up to an element $w \in \Gamma$:
\begin{equation}
p q \: = \: w q p.
\end{equation}
The matrices $p$ and $q$ are the holonomies of any bundle about
two cycles of the torus, lifted from $G/\Gamma$ to
$G$.  Put another way, these almost-commuting holonomies are the result
of lifting commuties holonomies in $G/\Gamma$ to pairs in $G$.
Next, we simultaneously diagonalize the adjoint action of $p$ and $q$
on the generators of the Lie algebra in the adjoint representation,
writing
\begin{eqnarray}
p T^{\alpha} p^{-1} & = & \omega_p^{\alpha} T^{\alpha},
\\
q T^{\alpha} q^{-1} & = & \omega_q^{\alpha} T^{\alpha},
\end{eqnarray}
where $\omega_{p,q}^{\alpha}$ are phases, which enter into the elliptic genus
computation.  These phases also appeared in the calculation of the four-dimensional Witten index \cite{Keurentjes:1999qf, Keurentjes:1999mv} .
Note that such a diagonalization is not
possible for every possible representation in which the $T^{\alpha}$ may appear;
in particular, for the diagonalization above to be possible, one needs
for the representation to be acted upon nontrivially\footnote{
A potentially useful reference is \cite{mo}, describing representations for
which such a diagonalization is possible.  For a representation in which
such a diagonalization is not possible, consider the case $G=SU(2)$, $\Gamma = 
{\mathbb Z}_2$, with $p$ and $q$ in the ${\bf 3}$ of $SU(2)$.  It is easy
to check that the resulting $3 \times 3$ matrices expressing the Lie algebra
simply cannot be diagonalized with respect to nontrivial $p$ and $q$.
} by the center detected
by $p$ and $q$.  Additionally the phases for the adjoint representation are sufficient to determine the phases for all representations
when the center of $G/\Gamma$ is trivial
since the adjoint is a tensor generator of the representation category \cite{DM:1982}.

If the phases $\omega_{p,q}$ are different from one, then, those `directions'
in the group are fixed.  If they are equal to one, on the other hand,
then the group is unconstrained in those directions, and so one must
integrate over corresponding Wilson lines, over the corresponding
moduli space of flat connections, to get the elliptic genus.

To the latter end, it can be shown that
\cite{Schweigert:1996tg, BFM2002}
\begin{equation}
{\cal M}_{G/\Gamma,w} \: = \: {\cal M}_{\tilde{G}(w),1}
\end{equation}
for some other group $\tilde{G}(w)$ that depends upon $G/\Gamma$ and $w$,
where ${\cal M}$ denotes the moduli space of flat connections.
Such groups $\tilde{G}(w)$ are listed in\footnote{
In addition, the paper \cite{fgl} relates
the moduli spaces $\tilde{G}(w)$ to $\tilde{G}(w=0)$ by Galois coverings.
}  \cite[section 5.4]{fm},
\cite[table 6]{Kac:1999gw}, and \cite[appendix A]{frat},
and we summarize their results in table~\ref{table:gw}.
Roughly speaking, we can think of the groups $\tilde{G}(w)$ as being
obtained by folding the affine Dynkin diagram according to the action of
$w \in \Gamma$.

\begin{table}
\begin{center}
\begin{tabular}{c|cc}
$G/\Gamma$ & $w$ & $\tilde{G}(w)$ \\ \hline
$A_{n-1} \sim SU(n)/{\mathbb Z}_n$ & $d$ & $SU(m)$, $m = gcd(n,d)$ \\ \hline
$B_n \sim {\rm Spin}(2n+1)/{\mathbb Z}_2$ & $1$ & $Sp(2n-2)$, Spin$(2n-1)$ \\ \hline
$C_{2n} \sim Sp(4n)/{\mathbb Z}_2$ & $1$ & $Sp(2n)$, Spin$(2n+1)$ \\
$C_{2n+1} \sim Sp(4n+2)/{\mathbb Z}_2$ & $1$ & $Sp(2n)$, Spin$(2n+1)$ \\ \hline
$D_{2n+1} \sim {\rm Spin}(4n+2)/{\mathbb Z}_4$ & $1$ & $Sp(2n-2)$, Spin$(2n-1)$ \\
 & $2$ & $Sp(4n-2)$, Spin$(4n-1)$  \\
 & $3$ & $Sp(2n-2)$, Spin$(2n-1)$ \\
$D_{2n} \sim {\rm Spin}(4n)/{\mathbb Z}_2 \times {\mathbb Z}_2$ & $(1,0)$ & $Sp(2n)$, Spin$(2n+1)$ \\
 & $(0,1)$ & $Sp(4n-4)$, Spin$(4n-3)$  \\
 & $(1,1)$ & $Sp(2n)$, Spin$(2n+1)$ \\ \hline
$E_6/{\mathbb Z}_3$ & $1$ & $G_2$ \\
  & $2$ & $G_2$ \\ \hline
$E_7/{\mathbb Z}_2$ & $1$ & $F_4$
\end{tabular}
\end{center}
\caption{List of groups $\tilde{G}(w)$ whose moduli space of flat
connections matches that of a moduli space of flat $G/\Gamma$ connections
with nontrivial characteristic class $w \in H^2(T^2,\Gamma)$.  In each case,
we assume $\Gamma$ is all of the center of simply-connected $G$, and not
a subgroup.  In $D_{2n}$, the $(0,1)$ 
indicates the ${\mathbb Z}_2$ whose quotient of Spin$(4n)$ is $SO(4n)$.  
Note that because the ranks and Weyl groups match, the moduli space of
flat Spin$(2k+1)$ connections matches that of flat $Sp(2k)$
connections.  This table summarizes results in \cite[section 5.4]{fm},
\cite[table 6]{Kac:1999gw}, and \cite[appendix A]{frat}.
\label{table:gw} }
\end{table}

To describe the moduli spaces ${\cal M}_{G/\Gamma,w = 0}$ more concretely
we recall some notions from the theory of compact Lie groups.
Let $T$ a maximal torus of $G/\Gamma$ \footnote{Not to be confused with the elliptic curve $T^2$.} with corresponding Cartan subalgebra $\fh.$
Let $Q$ be the root lattice, $P$ be the weight lattice, and $\Lambda_{char}$ be the character lattice of $G/\Gamma$.
Similarly, let $Q^{\vee}$ be the coroot lattice, $P^{\vee}$ be the coweight lattice, and $\Lambda^{\vee}_{char}$ be the co-character lattice.
Then the Cartan torus of $G/\Gamma$ can be identified with $\fh/2\pi { \Lambda}_{char}^{\vee}$.
The center of and fundamental groups of $G/\Gamma$ are
\begin{align}
Z(G/\Gamma) & \cong P^{\vee}/\Lambda^{\vee}_{char} \cong \Lambda_{char}/Q, \\
\pi_1(G/\Gamma) & \cong \Lambda^{\vee}_{char}/Q^{\vee} \cong P/\Lambda_{char}.
\end{align}
Let
\begin{equation}
\label{eq:moduli space}
\fM =\fh_\bC /({ \Lambda}_{char}^{\vee} + \tau {\Lambda}_{char}^{\vee}) \;,
\end{equation}
then the moduli space of flat $G/\Gamma$-connections on $T^2$ with $w = 0$ is 
\begin{equation}
{\cal M}_{G/\Gamma,w = 0} = \fM/W,
\end{equation}
where $W$ is the Weyl group of $G/\Gamma$.

For $G$ simply-connected the cocharacter lattice is equal to the coroot lattice.  In the opposite extreme of $G/\Gamma$
with trivial center, the cocharacter lattice is equal to the coweight lattice.  The relations between the cocharacter lattices mean that the moduli space
${\cal M}_{G, 1}$ is an order $|\Gamma \times \Gamma|$ cover of ${\cal M}_{G/\Gamma,w = 0}.$

The elliptic genus of a pure $G/\Gamma$ theory (with bundles of
vanishing characteristic class) is given by
\cite{Benini:2013xpa}
\begin{equation}
\label{eq:eg}
Z_{T^2}(\tau,z,w=0) = \frac1{|W|} \sum_{u_* \,\in\, \fM_\text{sing}^*} \JKres_{u=u_*}\!\big(\sQ(u_*),\eta \big) \;\; Z_\text{1-loop}(\tau,z, u)
\end{equation}
where $|W|$ is the order of the Weyl-group of $G.$\footnote{We omit the flavor holonomies $\xi$ since they are absent in pure theories.}
The Jeffrey--Kirwan residue operation \cite{JeffreyKirwan} $ \JKres_{u=u_*}\!\big(\sQ(u_*),\eta \big)$ assigns a residue to each pole of $Z_\text{1-loop}$ in $\fM_\text{sing}^*$
depending on the charge vectors $\sQ(u_*)$ responsible for the pole and a covector $\eta.$
The parameter $q = e^{2\pi i \tau}$ in $Z_\text{1-loop}$ specifies the complex structure of the torus $T^2$ and $y = e^{2\pi i z}$ is the fugacity for the left-moving $U(1)$ R-symmetry.
The coordinates $u_a$ on the moduli space $\fM$ can equivalently be described by the coordinates $x_a = e^{2\pi i u_a}.$
The contribution of a vector multiplet $V$ with gauge group $G/\Gamma$ to $Z_\text{1-loop}$ for the $w = 0$ characteristic class is
\begin{equation}
\label{1-loop vector}
Z_{V,G/\Gamma}(\tau,z,u) = \bigg( \frac{2\pi\eta(q)^3}{\theta_1(q,y^{-1})} \bigg)^{\rank G} \; \prod_{\alpha \,\in\, G} \frac{\theta_1(q, x^\alpha)}{\theta_1(q, y^{-1} x^\alpha)} \; \prod_{a=1}^{\rank G} \dd u_a\;.
\end{equation}
The product is over the roots $\alpha$ of the gauge group and $\eta(q)$ is the Dedekind eta function.

For bundles with non-trivial characteristic classes $w$, the contribution to $Z_{\text{1-loop}}$ is modified.
Using the eigenvalues $\omega_{p,q}^{\alpha}$, one can then
construct an elliptic genus for bundles of fixed
characteristic class $w$ as a product of ratios 
\begin{equation}
\frac{
\theta_1( \tau | v_{\alpha})
}{
\theta_1( \tau | -z + v_{\alpha})
},
\end{equation}
for nonzero $v_{\alpha}$,
where
\begin{equation}
v_{\alpha} \: = \: \ln \frac{\omega_p^{\alpha}}{2\pi i} \: + \:
\tau \ln \frac{\omega_q^{\alpha}}{2\pi i},
\end{equation}
and a residue integral of the form
\begin{equation}
\label{eq:EGw}
\bigg( \frac{2\pi\eta(q)^3}{\theta_1(q,y^{-1})} \bigg)^{\rank \tilde{G}(w)} \prod_{\alpha \,\in\, G} \frac{
\theta_1( \tau | v_{\alpha})
}{
\theta_1( \tau | -z + v_{\alpha})
} \;  \prod_{a=1}^{\rank \tilde{G}(w)} \dd u_a\;.
\end{equation}
for every vanishing $v$.  The resulting residue integral is computed
as a Jeffrey-Kirwan residue over (a cover of) the moduli space of those
flat connections preserving the holonomy.

This determines the elliptic genus (for fixed bundle characteristic
class $w$) up to
an overall normalization factor, which reflects residual gauge
transformations that preserve the holonomies.  For theories of the
form $SU(n)/{\mathbb Z}_n$, that normalization factor is computed
in e.g. \cite[section 2.2.1]{Kologlu:2016aev}.

So far we have described how one computes contributions to the
elliptic genus from bundles with different characteristic classes
$w \in H^2(T^2,\Gamma)$.  Finally, we will combine them, to form the elliptic genus
as a function of the discrete theta angle.  These different 
contributions are each weighted with potentially two different
phases.  First, there is a factor $\exp(i \theta \cdot w)$,
where $\theta \in \hat{\Gamma}$ is a choice of discrete theta angle.
Second, as studied in detail in \cite{Gu:2020ivl} and reviewed in
section~\ref{sect:rev:nsc},
there is a factor of the form
$\exp( i  w \cdot t )$,
where \cite[equ'n (2.7)]{Gu:2020ivl}
\begin{equation}
t_a \: = \: - \frac{\pi i}{|\Gamma|} \sum_{ \tilde{\mu}\: {\rm pos'}} 
\alpha^a_{\tilde{\mu}},
\end{equation}
and $w$ is encoded in $w_a$ so that
\begin{equation}
t \cdot w \: = \: \sum_a t_a w_a.
\end{equation}
Strictly speaking, the $t_a$ are not uniquely defined, as there are e.g.
branch cut ambiguities, 
but the phase factor above is well-defined, 
as discussed in detail in \cite{Gu:2020ivl}.
Put another way, the $t_a$ encode a constant shift, due to quantum
corrections, to the discrete theta angle $\theta$.

Thus, if we label the contribution to the elliptic genus of a pure
$G/\Gamma$ gauge theory in a sector with bundles of characteristic class $w$
by $Z( G/\Gamma, w)$, then the elliptic genus for a general characteristic
class has the form
\begin{equation}
Z(G/\Gamma, \theta) \: = \: \sum_w \exp(i w \cdot \theta)
\exp\left(i  w \cdot t\right) Z( G/\Gamma, w).
\end{equation}

In the next several sections we will carry out this program for
several low-rank examples. 
Specifically, we 
will apply the procedure above to derive elliptic genera for
$SU(2)/{\mathbb Z}_2$, $SU(3)/{\mathbb Z}_3$, $SO(4)$, Spin$(4)/({\mathbb Z}_2
\times {\mathbb Z}_2)$, $SO(5)$, and $Sp(6)/{\mathbb Z}_2$ gauge theories
with various discrete theta angles.  The special case of 
$SU(2)/{\mathbb Z}_2$ was previously discussed in
\cite[appendix A]{Kim:2017zis}; we recover their results through this
systematic method.  In each case, we will find that the
elliptic genus vanishes unless the discrete theta angle takes the
value listed in table~\ref{table:gw},
as expected \cite{Gu:2018fpm,Chen:2018wep,Gu:2020ivl}.
We will also see that the results are consistent with
decomposition \cite{Hellerman:2006zs,Sharpe:2014tca,Sharpe:2019ddn}.

Furthermore,
in each case we discuss, we will also find that the
contribution to the elliptic genus from bundles with characteristic
class $w \neq 0$ matches (up to a phase) the contribution from bundles
of characteristic class $w = 0$.  This is reminiscent of the fact
that elliptic genera are independent of deformations, and so one is
naturally led to wonder if there is a more elegant approach to
these computations that demonstrates that contributions to the
elliptic genus are (modulo an overall phase) independent of $w$.
For example, for sigma models on Calabi-Yau manifolds,
the scale $r$ of the Calabi-Yau is a marginal parameter, so as the
elliptic genus is an index, it is independent of $r$, and the resulting
elliptic genera are necessarily independent of worldsheet instanton
corrections \cite{Witten:1986bf,Witten:1987cg}.  In two-dimensional
gauge theories, on the other hand,
the gauge coupling is irrelevant\footnote{
We should be careful as terms such as `marginal' and `irrelevant'
are not well-defined away from fixed points of renormalization group
flow, but we are not aware of examples of two-dimensional (2,2)
supersymmetric gauge theories in which the gauge coupling flows in the IR
to a marginal operator.
}, so this argument does not apply.  In any event, we leave this question
for future work.

\section{Pure $SU(2)/{\mathbb Z}_2 = SO(3)$ gauge theories}
\label{sect:so3}

The elliptic genus of pure $SU(2)$ gauge theory is \cite{Benini:2013nda}
\begin{equation}
\frac12 \sum_{u_* \,\in\, \fM^+_\text{sing}}
\frac{i \eta(q)^3}{\theta_1(\tau|-z)}
\oint_{u_*}\dd u\,
\frac{\theta_1(\tau|2u)}{\theta_1(\tau|-z+2u)} \, 
\frac{\theta_1(\tau|-2u)}{\theta_1(\tau| -z - 2u)}, \\
\end{equation}
where the contributing poles are located at
\begin{equation}
\label{eq:SU2poles}
\fM^+_\text{sing}=\Big\{  \frac z2 \,,\,  
\frac{z+1}2 \,,\,  \frac{z+\tau}2 \,,\,  \frac{z+\tau+1}2 
\Big\} \;.
\end{equation}
Elliptic genera of pure $SO(3)$ gauge theories were computed
in \cite[appendix A]{Kim:2017zis}.
Briefly, the authors argued that the pure $SU(2)$
and the $SO(3)_-$ theories have the same elliptic genus,
given by
\begin{equation}  \label{eq:su2:eg}
\frac{ \theta_1( \tau | -z ) }{ \theta_1( \tau | -2z) }
\: = \:
\frac{1}{2}
\frac{ \theta_1( \tau | + 1/2 ) }{ \theta_1( \tau | -z + 1/2) }
\frac{ \theta_1( \tau | + \tau/2) }{ \theta_1( \tau | -z + \tau/2) }
\frac{ \theta_1( \tau | - (1 + \tau)/2) }{ \theta_1( \tau | -z - (1 + \tau)/2)},
\end{equation}
while the elliptic genus of the pure $SO(3)_+$ theory vanishes 
identically.  This is consistent with the results of \cite{Gu:2020ivl},
which argued that in pure $SO(3)$ gauge theories, only for the
nontrivial discrete theta angle are there supersymmetric vacua,
and supersymmetry is broken in the IR in $SO(3)_+$.  It is also
consistent with decomposition 
\cite{Hellerman:2006zs,Sharpe:2014tca,Sharpe:2019ddn}, which in this case
can be schematically expressed as
\begin{equation}
SU(2) \: = \: SO(3)_+ \: + \: SO(3)_-.
\end{equation}

In more detail, \cite[appendix A]{Kim:2017zis}
combined the contributions of the two distinct types of $SO(3)$ bundles.
The contribution to the $SO(3)$ elliptic genus from
bundles of vanishing characteristic class is obtained from
\begin{equation}
\frac12 \sum_{u_* \,\in\, \fM^+_\text{sing}}
\frac{i \eta(q)^3}{\theta_1(\tau|-z)}
\oint_{u_*}\dd u\,
\frac{\theta_1(\tau|u)}{\theta_1(\tau|-z+u)} \, 
\frac{\theta_1(\tau|-u)}{\theta_1(\tau| -z - u)}, \\
\end{equation}
with a single contributing pole located at $\fM^+_\text{sing}= z/2.$
This results in
\begin{equation}
Z(SO(3)_0) \: = \: \frac{1}{2} \frac{
\theta_1( \tau | -z ) }{ \theta_1( \tau | -2 z) },
\end{equation}
which is the $SU(2)$ elliptic genus up to a factor of $1/|\Gamma| = 1/2$.
As explained in section~\ref{sect:strategy},
this factor arises from the differing character lattices of the $SU(2)$ and $SO(3)$ groups. 
Note that all four poles in equation~\ref{eq:SU2poles} contribute equally to the $SU(2)$ elliptic genus, but there is only
one pole for the $SO(3)$ elliptic genus.
Since there are only $1/|\Gamma \times \Gamma| = 1/4$ as many poles, but each pole 
has a Jacobian contribution of 
$|\Gamma|$ relative to the $SU(2)$ poles, we arrive at the previously claimed factor of $|\Gamma|/|\Gamma \times \Gamma| = 1/|\Gamma| = 1/2.$

The contribution from bundles of nonzero characteristic class is
\begin{equation}
Z( SO(3)_1 ) \: = \: - \frac{1}{2} \frac{
\theta_1( \tau | -z ) }{ \theta_1( \tau | -2 z) }.
\end{equation}
For a discrete theta angle $\theta \in \{ 0, \pi \}$,
the possible $SO(3)$ elliptic genera are
\begin{eqnarray}
Z(SO(3)) & = & Z(SO(3)_0) \: + \: \exp(i \theta) Z(SO(3)_1),
\\
& = & \frac{1}{2}  \frac{
\theta_1( \tau | -z ) }{ \theta_1( \tau | -2 z) }
\left( 1 - \exp(i \theta) \right).
\end{eqnarray}
When $\theta = 0$, this vanishes, and when $\theta = \pi$,
this is nonzero and matches $Z(SU(2))$.

For later use, the elliptic genus, given up to numerical factors
we will describe later, is
\begin{equation}
\frac{ \theta_1( \tau | + 1/2 ) }{ \theta_1( \tau | -z + 1/2) }
\frac{ \theta_1( \tau | + \tau/2) }{ \theta_1( \tau | -z + \tau/2) }
\frac{ \theta_1( \tau | - (1 + \tau)/2) }{ \theta_1( \tau | -z - (1 + \tau)/2)}
\end{equation}
can be derived 
directly from thinking about the contribution of
$Z_{{\rm 1-loop}}$ in the sector with $w_2 \neq 0$, 
in the notation of \cite{Benini:2013nda}.  Briefly, for $w_2 \neq 0$,
the moduli space of flat connections is a point, so that one does
not integrate over a space of $u$'s.  Instead, the $u$'s are fixed, with
holonomies about the $T^2$ which can be taken to be
\begin{equation}
{\rm diag}(-1, -1, +1), \: \: \:
{\rm diag}(+1,-1,-1).
\end{equation}

An $SO(3)$ bundle with these holonomies cannot be
lifted to an $SU(2)$ bundle.
A heuristic way to see this is to observe that the lifts of the holonomies to $SU(2)$ are given in equation 
~\ref{eq:Pauli} and they anticommute.
We can also see this more formally by computing the second Stiefel-Whitney
class $w_2$, which gives the obstruction to lifting, in this case,
an $SO(3)$ bundle to an $SU(2)$ bundle.  With the holonomies above,
we can describe this bundle as
\begin{equation}
L_1 \oplus L_2 \oplus L_3,
\end{equation}
where $L_1$ and $L_2$ each have nontrivial monodromy about a single
$S^1$ on $T^2$, and $L_3 = L_1 \otimes L_2$.  Thus, for example,
\begin{equation}
w(L_1) \: = \: 1 + J_1, \: \: \:
w(L_2) \: = \: 1 + J_2, \: \: \:
w(L_3) \: = \: 1 + J_1 + J_2,
\end{equation}
where $J_{1}$, $J_2$ generate $H^1(T^2,{\mathbb Z}_2) = ({\mathbb Z}_2)^2$,
and in this case give $w_1$ of $L_1$, $L_2$, respectively.
Thus,
\begin{equation}
w(L_1 \oplus L_2 \oplus L_3) \: = \: w(L_1) w(L_2) w(L_3) \: = \:
1 + J_1 J_2 + \cdots,
\end{equation}
hence
\begin{equation}
w_2(L_1 \oplus L_2 \oplus L_3) \: = \: J_1 J_2,
\end{equation}
and in particular is nonzero.  Thus, indeed, this $SO(3)$ bundle has
nonzero $w_2$, and can not be lifted to an $SU(2)$ bundle.

Returning to the computation of the elliptic genus for the pure
$SO(3)$ gauge theory in a sector in which $w_2 \neq 0$, in terms of 
holonomies encoded in the parameter $u$, it can be written
\begin{equation}
\prod_{{\rm roots} \, \alpha} \frac{ \theta_1( \tau | \alpha \cdot u )}{
\theta_1( \tau | -z + \alpha \cdot u) }
\: = \:
\frac{ \theta_1( \tau | u) }{ \theta_1( \tau | -z + u) }
\frac{ \theta_1( \tau | 0) }{ \theta_1( \tau | -z) }
\frac{ \theta_1( \tau | -u) }{ \theta_1( \tau | -z - u) },
\end{equation}
corresponding to the three generators of the Lie algebra of $SO(3)$.
The three boundary conditions correspond to values of $u$ as follows:
\begin{center}
\begin{tabular}{c|c}
$u$ & $(U_1,U_2)$ \\ \hline
$0$ & $(+1,+1)$ \\
$1/2$ & $(-1,+1)$ \\
$\tau/2$ & $(+1,-1)$ \\
$(1+\tau)/2$ & $(-1,-1)$
\end{tabular}
\end{center}
Plugging in the single holonomy, we find that the elliptic genus for the
$w_2 \neq 0$ sector of the pure $SO(3)$ gauge theory is
proportional to
\begin{equation} \label{eq:so3:wneq0}
\frac{ \theta_1( \tau | + 1/2 ) }{ \theta_1( \tau | -z + 1/2) }
\frac{ \theta_1( \tau | + \tau/2) }{ \theta_1( \tau | -z + \tau/2) }
\frac{ \theta_1( \tau | - (1 + \tau)/2) }{ \theta_1( \tau | -z - (1 + \tau)/2)},
\end{equation}
confirming the results of \cite[appendix A]{Kim:2017zis} up to numerical
factors we will describe momentarily.

So far, we have discussed known results for $SU(2)$ elliptic genera,
and also used a trick to compute the $SO(3)$ elliptic genus
in a sector where the characteristic class is nontrivial.  Let us now
repeat the computation systematically using the method of
section~\ref{sect:strategy}, which we will apply to other examples.

Following the method of section~\ref{sect:strategy}, we compute
the contribution to the elliptic genus from $SO(3)$ bundles of
vanishing characteristic (Stiefel-Whitney) class $w_2$.  As discussed
there, the contribution in this sector is the same as that of a pure
$SU(2)$ theory, albeit with a constant factor of $1/|\Gamma| = 1/2$ from section~\ref{sect:strategy}:
\begin{equation}
Z( SO(3), w_2=0) \: = \: \frac{1}{2}
\frac{ \theta_1( \tau | -z) }{ \theta_1( \tau | -2z) }.
\end{equation}

Next, we compute the contribution from $SO(3)$ bundles of
nontrivial characteristic class.  As in section~\ref{sect:strategy},
we define this sector through holonomies lifted to $SU(2)$,
where they anticommute.
Specifically, consider the $SU(2)$ matrices
\begin{equation}
\label{eq:Pauli}
p \: = \: \left[ \begin{array}{cc} 0 & 1 \\ -1 & 0 \end{array} \right],
\: \: \:
q \: = \: \left[ \begin{array}{cc} i & 0 \\ 0 & -i \end{array} \right].
\end{equation}
It is easy to verify that $pq = - qp$.  Viewing $p$ and $q$ as holonomies, 
they define a flat $SU(2)/{\mathbb Z}_2 = SO(3)$ 
bundle with nontrivial characteristic class.
Under the adjoint action of $p$ and $q$, the Pauli sigma matrices are
diagonal:
\begin{equation}
p \sigma_1 p^{-1} \: = \: - \sigma_1, 
\: \: \:
p \sigma_2 p^{-1} \: = \: + \sigma_2,
\: \: \:
p \sigma_3 p^{-1} \: = \: - \sigma_3,
\end{equation}
\begin{equation}
q \sigma_1 q^{-1} \: = \: - \sigma_1,
\: \: \:
q \sigma_2 q^{-1} \: = \: - \sigma_2,
\: \: \:
q \sigma_3 q^{-1} \: = \: + \sigma_3.
\end{equation}

From table~\ref{table:gw}, we see that the moduli
space of flat $SO(3)$ connections with nontrivial characteristic class
is a point.
We compute the contribution to the elliptic genus for this nontrivial characteristic class
by applying equation~\ref{eq:EGw} with the phases listed above
to get the the product of theta functions in equation~\ref{eq:so3:wneq0} up to a constant factor.

Finally, to derive the elliptic genus for bundles of nonzero second
Stiefel-Whitney class, we need to add a suitable numerical factor,
corresponding to dividing out by the number of residual gauge transformations
which preserve the holonomies.  From  \cite[section 2.2.1]{Kologlu:2016aev}
for this case, we multiply the theta function product~(\ref{eq:so3:wneq0})
by a factor of $1/|W|$, where $W = {\mathbb Z}_2 \times {\mathbb Z}_2$.
Thus, we have that
\begin{eqnarray}
Z(SO(3), w_2 \neq 0) & = &
\frac{1}{4} 
\frac{ \theta_1( \tau | + 1/2 ) }{ \theta_1( \tau | -z + 1/2) }
\frac{ \theta_1( \tau | + \tau/2) }{ \theta_1( \tau | -z + \tau/2) }
\frac{ \theta_1( \tau | - (1 + \tau)/2) }{ \theta_1( \tau | -z - (1 + \tau)/2)},
\nonumber \\
& = &
\frac{1}{2} \frac{
\theta_1( \tau | -z)
}{
\theta_1( \tau | -2z) }.
\end{eqnarray}

Now, let us assemble these contributions.
For a discrete theta angle $\theta$,
\begin{equation}
Z( SO(3), \theta) \: = \: 
Z( SO(3), w_2 = 0) \: + \: \exp(i w \cdot t) \exp(i w \cdot \theta)
Z( SO(3), w_2 \neq 0).
\end{equation}
As computed in \cite[section 3.1]{Gu:2020ivl}, $t = - \pi i$,
hence
\begin{equation}
\exp(i w \cdot t) \: = \: -1,
\end{equation}
and trivially $\exp(i w \cdot \theta) = \exp(i \theta)$,
hence
\begin{eqnarray}
Z(SO(3), \theta) & = &
Z( SO(3), w_2=0) \: - \: \exp(i \theta) Z( SO(3), w_2 \neq 0),
\\
& = &
\frac{1}{2}
\frac{ \theta_1( \tau | -z) }{ \theta_1( \tau | -2z) }
\left( 1 - \exp(i \theta) \right),
\end{eqnarray}
which duplicates the $SO(3)$ elliptic genus as a function of $\theta$
computed in \cite[appendix A]{Kim:2017zis}.

\section{Pure $SU(3)/{\mathbb Z}_3$ gauge theories}
\label{sect:su3}

In this section, we apply the method of section~\ref{sect:strategy}
to compute the elliptic genus of a pure supersymmetric $SU(3)/{\mathbb Z}_3$
as a function of the discrete theta angle.
First, for a vanishing characteristic class,
from equation~(\ref{eq:factor}),
the elliptic genus of the pure $SU(3)/{\mathbb Z}_3$ gauge theory
is the same as the elliptic genus of the pure $SU(3)$ gauge theory, up to a 
factor of $1/|\Gamma|$:
\begin{equation}
Z( SU(3)/{\mathbb Z}_3, w = 0 ) \: = \:
\frac{1}{3} Z( SU(3) ) \: = \: 
\frac{1}{3} \frac{ \theta_1(\tau | -z) }{ \theta_1(\tau | -3z) }.
\end{equation}

Next, we consider the elliptic genus of a pure $SU(3)/{\mathbb Z}_3$
gauge theory with a nontrivial characteristic class.
We can describe an $SU(3)/{\mathbb Z}_3$ 
bundle with nonzero $w \in H^2(T^2, {\mathbb Z}_3)$
as two holonomies $p$ and $q$ in $SU(3)$ such that
\begin{equation}
p q \: = \: w q p
\end{equation}
for $w = \exp(2 \pi i k/3)$ with $k = \pm 1.$
To that end,
consider the $SU(3)$ matrices
\begin{equation}
p \: = \: \left[ \begin{array}{ccc} 
w & 0 & 0 \\ 0 & 1 & 0 \\ 0 & 0 & w^{-1} \end{array} \right],
\: \: \:
q \: = \: \left[ \begin{array}{ccc}
 0 & 1 & 0 \\ 0 & 0 & 1 \\
1 & 0 & 0 \end{array} \right],
\end{equation}
then, using $w^3 = 1$,
one can verify that
\begin{equation}
p q \: = \: w q p.
\end{equation}

Taking linear combinations of the Lie algebra generators
$\lambda_a$ (in the three-dimensional adjoint representation) to solve
\begin{equation}
p \lambda_a p^{-1} \: = \: \omega_p^a \lambda_a,
\: \: \:
q \lambda_a q^{-1} \: = \: \omega_q^a \lambda_a,
\end{equation}
we find that 
\begin{equation}
(\omega_p, \omega_q) \: \in \: \{
(1, w), (1, w^2), (w^2, 1), (w^2, w^2), (w^2, w),
(w,1), (w,w^2), (w,w) \}.
\end{equation}
In particular, the dimension of this component of the moduli space
of flat $SU(3)/{\mathbb Z}_3$ connections is zero, as can be
confirmed from table~\ref{table:gw}.

Using equation~\ref{eq:EGw} with these phases, we find that
the elliptic genus of the pure $SU(3)/{\mathbb Z}_3$ gauge theory 
with nontrivial bundle is
\begin{eqnarray}
\lefteqn{
\frac{1}{|W|}
\frac{ \theta_1( \tau \, | \, \tau k/3) }{ \theta_1( \tau \, | \, -z + \tau k/3) }
\frac{ \theta_1( \tau \, | \, - \tau k/3) }{ \theta_1( \tau \, | \, -z - \tau k/3) }
\frac{ \theta_1( \tau \, | \, -k/3) }{ \theta_1( \tau \, | \, -z - k/3) }
\frac{ \theta_1( \tau \, | \, -k/3 - \tau k/3) }{
       \theta_1( \tau \, | \, -z  -k/3 - \tau k/3) }
} \nonumber \\
& & \cdot
\frac{ \theta_1( \tau \, | \, -k/3 + \tau k/3) }{ 
       \theta_1( \tau \, | \, -z  -k/3 + \tau k/3) }
\frac{ \theta_1( \tau \, | \, k/3 ) }{ \theta_1( \tau \, | \, -z + k/3) }
\frac{ \theta_1( \tau \, | \, k/3 - \tau k/3 ) }{
       \theta_1( \tau \, | \, -z + k/3 - \tau k/3) }
\nonumber \\
& & \cdot
\frac{ \theta_1( \tau \, | \, k/3 + \tau k/3 ) }{
       \theta_1( \tau \, | \, -z + k/3 + \tau k/3) },
\end{eqnarray}
where $W$ is the unbroken gauge symmetry of the pair
$(p,q)$, which for this case is 
\cite[section 2.2.1]{Kologlu:2016aev}
$W = {\mathbb Z}_3 \times
{\mathbb Z}_3$, hence $|W| = 9$.
Recall that $w = \exp(2 \pi i k/3)$ for $k=\pm 1$ (corresponding to the two
nontrivial possible values of the characteristic class), 
Note that this expression is symmetric under $k \leftrightarrow -k$.

For $k=1$, the product above can be written more succinctly as
\begin{equation}
\frac{1}{9}
\prod_{j,\ell=-1}^{1}  \left[ 
\frac{ \theta_1( \tau | j/3 + \ell \tau/3 ) }{
\theta_1( \tau | j/3 + \ell \tau/3 - z ) }
\frac{ \theta_1( \tau | -z ) }{ \theta_1( \tau | 0 ) }
\right],
\end{equation}
where in the product one should omit the case
$j=k=0$.  One can show that\footnote{A careful reader will observe
that if we had instead chosen $k = 1, 2$, we would have crossed a branch
cut, which can generate factors such as $y^3$.  We note that fact here,
but it will not play a role in our further computations.}
\begin{equation}
\prod_{j,\ell=-1}^{1}   \left[ 
\frac{ \theta_1( \tau | j/3 + \ell \tau/3 ) }{
\theta_1( \tau | j/3 + \ell \tau/3 - z ) }
\frac{ \theta_1( \tau | -z ) }{ \theta_1( \tau | 0 ) }
\right]
\: = \:
3 \,
\frac{
\theta_1(\tau | - z) }{ \theta_1(\tau | -2 z) }
\frac{ 
\theta_1( \tau | -2 z) }{ \theta_1(\tau | -3 z) },
\end{equation}
where $y = \exp(2 \pi i z)$.

Now, let us assemble these pieces to build the elliptic genus of
the pure $SU(3)/{\mathbb Z}_3$ theory with discrete theta angle
$\theta \in \{0, 2 \pi/3, 4 \pi/3\}$.  
From \cite[section 3.2]{Gu:2020ivl}, the quantum correction 
is given by
\begin{equation}
t_a \: = \: \frac{2 \pi i}{3} m_a,
\end{equation}
where
\begin{equation}
\sum_a m_a \: \equiv \: 0 \mod 3.
\end{equation}
Without loss of generality, we can choose $m_1 = 0 = m_2$,
hence the phase factor 
\begin{equation}
\exp(i w \cdot t) = +1,
\end{equation}
and so the elliptic genus can be written as a function of $\theta \in
\{ 0, 2 \pi /3, 4\pi /3 \}$ as
\begin{eqnarray}
Z( SU(3)/{\mathbb Z}_3, \theta) & = &
Z( SU(3)/{\mathbb Z}_3, w=0) \: + \:
\exp(i \theta) Z( SU(3)/{\mathbb Z}_3, w=1)
\nonumber \\
& & \: + \:
\exp(-i\theta) Z( SU(3)/{\mathbb Z}_3, w=2),
\\
& = &
\frac{1}{3} \frac{ \theta_1( \tau | -z) }{ \theta_1( \tau | -3z)}
\left( 1 \: + \:  \exp(i \theta) \: + \: \exp(- i \theta)
\right).
\end{eqnarray}
As a consistency check, the reader should note that for $\theta \neq 0$,
the expression above for the elliptic genus vanishes, 
whereas for $\theta=0$, it matches that of the pure $SU(3)$ gauge
theory.  This is 
consistent with the computation in
\cite[section 3.2]{Gu:2020ivl} that supersymmetry is only unbroken
in a supersymmetric pure $SU(3)/{\mathbb Z}_3$ gauge theory when
$\theta=0$.

Furthermore, 
\begin{equation}
\sum_{\theta = 0, \pm 2 \pi/3} 
\frac{1}{3} \frac{ \theta_1( \tau | -z) }{ \theta_1( \tau | -3z)}
\left( 1 \: + \:  \exp(i \theta) \: + \: \exp(- i \theta)
\right) \: = \:
\frac{ \theta_1( \tau | -z) }{ \theta_1( \tau | -3z)} \: = \: Z( SU(3) ).
\end{equation}
This matches the prediction of decomposition
\cite{Hellerman:2006zs,Sharpe:2014tca,Sharpe:2019ddn},
which in this case schematically says that
\begin{equation}
SU(3) \: = \: \left( SU(3)/{\mathbb Z}_3 \right)_{\theta = 0}
\: + \:
 \left( SU(3)/{\mathbb Z}_3 \right)_{\theta = 2 \pi/3}
\: + \:
 \left( SU(3)/{\mathbb Z}_3 \right)_{\theta = 4 \pi/3}.
\end{equation}

\section{Pure $SO(4)$ gauge theories}
\label{sect:so4}

Let us now turn to the elliptic genera of pure $SO(4)$ gauge theories.
These can be derived from the results above for pure $SO(3)$ gauge
theories.

First, consider a pure $SO(4)$ theory in the sector in which
$w_2$ vanishes (so that all bundles can be lifted to Spin$(4)$ bundles).
Now, Spin$(4) = SU(2) \times SU(2)$, so the elliptic genus in this
sector is the product of elliptic genera corresponding to two pure
$SU(2)$ gauge theories.  Thus, as explained in section~\ref{sect:strategy}, 
the elliptic
genus of a pure $SO(4)$ gauge theory in a sector with $w_2=0$ is
\begin{equation}
\frac{1}{2}
\left( 
\frac{ \theta_1( \tau | -z ) }{ \theta_1( \tau | -2z) }
\right)^2,
\end{equation}
taking into account the constant factor of $1/|\Gamma|$ from section~\ref{sect:strategy}.
This is consistent with the prediction~(\ref{eq:ellgen-pred}) since
there are two Casimirs each of the form Tr $\Sigma^2$.

Now, let us turn to the sector in which $w_2 \neq 0$.
Here, we can apply the same analysis as in the case of the analogous
$SO(3)$ sectors.  A set of holonomies describing such $SO(4)$ bundles are
given by
\begin{equation}
{\rm diag}(+1, -1, -1, +1), \: \: \:
{\rm diag}(+1, +1, -1, -1).
\end{equation}
It is straightforward to check that these holonomies describe an $SO(4)$ bundle
with nonzero $w_2$, and from table~\ref{table:gw}, the moduli space
of flat $SO(4)$ connections with nonzero $w_2$ is a point.  
These holonomies emerge as a special case of the results in
\cite[equ'n (3.2)]{Kim:2014dza}.
We can think of these holonomies as describing transformations under one
of the two factors in $SO(4) = (SU(2) \times SU(2))/{\mathbb Z}_2$.
Now, the nonzero roots of $SO(4)$ can be expressed as
\begin{equation}
\pm u_1 \pm u_2,
\end{equation}
where $u_{1,2}$ couple to Cartan holonomies.  If one of the two $SU(2)$ factors
has trivial holonomy, then we can set $u_1 = 0$, in which case, these
roots become two copies of the roots of $SO(3)$.  Using previous
results for $SO(3)$ holonomies and elliptic genera, we immediately have
that the $SO(4)$ elliptic genus for $w_2 \neq 0$ is proportional to
\begin{equation}
\left[ \frac{ \theta_1( \tau | + 1/2 ) }{ \theta_1( \tau | -z + 1/2) }
\frac{ \theta_1( \tau | + \tau/2) }{ \theta_1( \tau | -z + \tau/2) }
\frac{ \theta_1( \tau | - (1 + \tau)/2) }{ \theta_1( \tau | -z - (1 + \tau)/2)}
\right]^2
\: = \: 
\left[ 2 
\frac{ \theta_1(\tau | -z) }{ \theta_1( \tau | -2z) }
\right]^2,
\end{equation}
which from equation~(\ref{eq:su2:eg}) is proportional to the elliptic genus
for pure $SO(4)$ gauge theories with vanishing $w_2$.

Now, let us assemble these contributions.  In principle, for
discrete theta angle $\theta \in \{0, \pi\}$,
\begin{equation}
Z(SO(4), \theta) \: = \: Z( SO(4), w=0) + \exp(i w \cdot t) 
\exp(i w \cdot \theta) Z(SO(4), w\neq 0).
\end{equation}
As computed in \cite{Gu:2020ivl}, $t_a = i \pi m_a$ where
\begin{equation}
\sum_a m_a \: \equiv \: 1 \mod 2,
\end{equation}
hence
\begin{equation}
\exp(i w \cdot t) \: = \: -1.
\end{equation}
Thus, the elliptic genus is given by
\begin{equation}
Z(SO(4), \theta) \: = \:
\frac{1}{2} \left( \frac{ \theta_1(\tau | -z) }{ \theta_1( \tau | -2z) }
\right)^2 \left( 1 \: - \: \exp(i \theta) \right).
\end{equation}

As a consistency check, note that $Z(SO(4), \theta)$ vanishes for
$\theta = 0$, which is consistent with the result \cite[section 13.1]{Gu:2018fpm}
that supersymmetry is broken in this theory for $\theta = 0$.  

As another consistency check, note that
\begin{equation}
\sum_{\theta = 0, \pi} 
\frac{1}{2} \left( \frac{ \theta_1(\tau | -z) }{ \theta_1( \tau | -2z) }
\right)^2 \left( 1 \: - \:  \exp(i \theta) \right)
\: = \: 
\left( \frac{ \theta_1(\tau | -z) }{ \theta_1( \tau | -2z) }
\right)^2,
\end{equation}
the elliptic genus of the pure Spin$(4)$ theory.
This confirms
the prediction of decomposition 
\cite{Hellerman:2006zs,Sharpe:2014tca,Sharpe:2019ddn} in this case,
which schematically says
\begin{equation}
{\rm Spin}(4) \: = \: SO(4)_{\theta = 0} \: + \:
SO(4)_{\theta = \pi}.
\end{equation}

\section{Pure Spin$(4)/({\mathbb Z}_2 \times {\mathbb Z}_2)$ gauge theories}
\label{sect:spin4}

The group Spin$(4) = SU(2) \times SU(2)$, so the analysis of this
group will be closely related to the analysis of $SU(2)$.
We can describe the Lie algebra of Spin$(4)$ in terms of block-diagonal
matrices 
and we can describe sectors with nontrivial characteristic classes
by taking holonomies to be of the form
\begin{equation}
{\rm diag}(p, 1), \: \: \: {\rm diag}(q,1)
\end{equation}
for one ${\mathbb Z}_2$ and
\begin{equation}
{\rm diag}(1,p), \: \: \: {\rm diag}(1,q)
\end{equation}
for the other ${\mathbb Z}_2$.  Proceeding in a simple generalization of
the analysis for a single copy of $SU(2)$, we find results for
elliptic genera as follows:
\begin{enumerate}
\item Vanishing characteristic class.  In this case, the elliptic genus
is a product of two copies of the $SU(2)$ elliptic genus (divided by a factor
of $| {\mathbb Z}_2 \times {\mathbb Z}_2 | = 4$):
\begin{equation}
\frac{1}{4} \left( Z( SU(2) ) \right)^2
\: = \: \frac{1}{4} \left( 
\frac{ \theta_1( \tau | -z ) }{ \theta_1( \tau | -2 z) } 
\right)^2.
\end{equation}
\item Nontrivial characteristic class in one ${\mathbb Z}_2$.  Here,
if we let $Z( SO(3)_1 )$ denote the elliptic genus of a single
$SO(3)$ theory with nontrivial characteristic class, then the elliptic
genus is given by
\begin{equation}
\frac{1}{2} Z( SU(2) ) Z( SO(3)_1 )
\: = \: \frac{1}{4}
\left( 
\frac{ \theta_1( \tau | -z ) }{ \theta_1( \tau | -2 z) } 
\right)^2
\end{equation}
(up to a phase).
\item Nontrivial characteristic classes in both ${\mathbb Z}_2$'s.
Here, the elliptic genus is given by 
\begin{equation}
\left( Z( SO(3)_1 ) \right)^2
\: = \:
\left( 
\frac{1}{2} 
\frac{ \theta_1( \tau | -z ) }{ \theta_1( \tau | -2 z) } 
\right)^2
\end{equation}
(up to a phase).
\end{enumerate}

In the expressions above, we have used that
\begin{equation}
Z( SU(2) ) \: = \: \frac{
\theta_1( \tau | - z) }{ \theta_1( \tau | -2 z) }
\end{equation}
and
\begin{equation}
Z( SO(3)_1 ) \: = \: \frac{1}{2} 
\frac{
\theta_1( \tau | - z) }{ \theta_1( \tau | -2 z) }
\end{equation}
up to a phase,
matching 
\cite{Kim:2017zis}.

Now, let us assemble these results.  In principle, a sector of bundles
of nontrivial characteristic class should be weighted by factors
$\exp(i w \cdot t)$ and
$\exp(i \theta)$, for $\theta$ a discrete theta angle, and using
results 
in \cite{Gu:2020ivl}, one can derive both phases for each sector.
However, in this case there is a faster way,
as the gauge group can equivalently be written as $SO(3) \times SO(3)$,
so we can reuse the results of \cite[appendix A]{Kim:2017zis} 
to immediately write the elliptic genus of a pure
Spin$(4)/{\mathbb Z}_2 \times {\mathbb Z}_2$ gauge theory with
discrete theta angles $(\theta_1, \theta_2)$, $\theta_i \in \{ 0, \pi\}$
as
\begin{equation}
\left[
\frac{ \theta_1( \tau | -z ) }{ \theta_1(\tau | -2z) }
\right]^2 
\left( 
\frac{ 1 - \exp(i \theta_1) }{2} \right)
\left( 
\frac{ 1 - \exp(i \theta_2) }{2} \right).
\end{equation}

In particular, note that
\begin{equation}
\sum_{ \theta_1, \theta_2 \in \{0, \pi\} }
\left[
\frac{ \theta_1( \tau | -z ) }{ \theta_1(\tau | -2z) }
\right]^2 
\left( 
\frac{ 1 - \exp(i \theta_1) }{2} \right)
\left( 
\frac{ 1 - \exp(i \theta_2) }{2} \right)
\: = \: 
\left[
\frac{ \theta_1( \tau | -z ) }{ \theta_1(\tau | -2z) }
\right]^2,
\end{equation}
and so we see that the elliptic genus of the pure Spin$(4)$ theory
matches that of the sum of the elliptic genera of pure
Spin$(4)/{\mathbb Z}_2 \times {\mathbb Z}_2$
theories with the various possible discrete theta angles,
as expected from decomposition 
\cite{Hellerman:2006zs,Sharpe:2014tca,Sharpe:2019ddn}
of two-dimensional theories with a
$B( {\mathbb Z}_2 \times {\mathbb Z}_2)$ symmetry.

\section{Pure $SO(5)$ gauge theories}
\label{sect:so5}

Now, let us turn to elliptic genera for pure $SO(5)$ gauge theories.
From equation~(\ref{eq:ellgen-pred}) and the fact that there are
two operators, tr $\Sigma^2$ and tr $\Sigma^4$, of R-charges $4$ and $8$,
one expects that the elliptic genus of the pure Spin$(5)$ theory
and that of a pure $SO(5)$ theory for one value of the discrete
theta angle is 
\begin{equation}
\frac{ \theta_1( \tau | - z) }{ \theta_1( \tau | -2 z ) }
\frac{ \theta_1( \tau | -3 z) }{ \theta_1( \tau | -4 z) },
\end{equation}
as discussed in section~\ref{sect:rev}.  This will also be derived by a direct residue computation in \cite{Eager}.

For bundles with vanishing $w_2,$
from equation~(\ref{eq:factor}),
the contribution to the elliptic genus of the pure $SO(5)$ gauge theory is $1/2$ of the elliptic genus
of the pure Spin$(5)$ theory
\begin{equation}  \label{eq:so5:w0}
\frac{1}{2}
\frac{ \theta_1( \tau | - z) }{ \theta_1( \tau | -2 z ) }
\frac{ \theta_1( \tau | -3 z) }{ \theta_1( \tau | -4 z) },
\end{equation}
as discussed in section~\ref{sect:strategy}.

Next, let us consider the case of nonzero $w_2$, which we analyze following
the pattern of section~\ref{sect:strategy}.
Following \cite[equ'n (3.3)]{Kim:2014dza}, we can express the holonomies
$p$, $q$ in the form
\begin{equation}
p \: = \:
{\rm diag}\left( \exp(2 \pi i \lambda_1 \sigma_2 ), -1, -1, +1 \right),
\: \: \:
q \: = \:
{\rm diag}\left( \exp(2 \pi i \lambda_2 \sigma_2), +1, -1, -1 \right).
\end{equation}
Since we have already descended to $SO(5)$ matrices, and are not working in
Spin$(5)$, these matrices commute.
Then, we diagonalize, finding a basis $T^{\alpha}$ of the Lie algebra such that
\begin{equation}
p T^{\alpha} p^{-1} \: = \: \omega_p^{\alpha} T^{\alpha},
\: \: \:
q T^{\alpha} q^{-1} \: = \: \omega_q^{\alpha} T^{\alpha}.
\end{equation}
Doing so, we find the eigenvalues $\omega_{p,q}^{\alpha}$, which we list
in table~\ref{table:so5}.  In each case, the $\theta$ argument is
computed as
\begin{equation}
\frac{\ln \omega_p^{\alpha}}{2 \pi i} \: + \: \tau \frac{\ln \omega_q^{\alpha}}{2 \pi i},
\end{equation}
and $u = \lambda_1 + \tau \lambda_2$.
The number of eigenvalues $(\omega_p,\omega_q) = (1,1)$ gives the dimension
of the residue integral, as it reflects moduli of flat connections that
are not constrained by the holonomies $p$, $q$.

\begin{table}
\begin{center}
\begin{tabular}{cc|c}
$\omega_p$ & $\omega_q$ & $\theta$ argument \\ \hline
$-1$ & $-1$ & $- (1 + \tau)/2$ \\
$-1$ & $+1$ & $1/2$ \\
$+1$ & $-1$ & $\tau/2$ \\
$+1$ & $+1$ & $0$ \\
$- \exp(2 \pi i \lambda_1)$ & $- \exp(2 \pi i \lambda_2)$ & 
$- (1 + \tau)/2 + u$ \\
$+ \exp(- 2 \pi i \lambda_1)$ & $- \exp(-2 \pi i \lambda_2)$ &
$\tau/2 - u$ \\
$-\exp(-2 \pi i \lambda_1)$ & $-\exp(-2 \pi i \lambda_2)$ &
$- (1+\tau)/2 - u$ \\
$\exp(2 \pi i \lambda_1)$ & $- \exp(2 \pi i \lambda_2)$ &
$\tau/2 + u$ \\
$-\exp(2 \pi i \lambda_1)$ & $+ \exp(2 \pi i \lambda_2)$ &
$1/2 + u$ \\
$- \exp(-2 \pi i \lambda_1)$ & $+\exp(- 2 \pi i \lambda_2)$ &
$1/2 - u$ 
\end{tabular}
\end{center}
\caption{List of eigenvalues of $SO(5)$ under the adjoint action of $p$,
$q$. \label{table:so5}}
\end{table}

Alternatively, one could think of table~\ref{table:so5}
in terms
of a (maximal-rank) $SO(2) \times SO(3)$ subgroup of $SO(5)$.
The weights of the nonzero roots of $SO(5)$ are
\begin{equation} \label{eq:so5weights}
\alpha \cdot u \: \in \: \left\{ \pm u_1 \pm u_2, \pm u_1, \pm u_2 \right\},
\end{equation}
In principle, for nonzero holonomies,
the product over roots is of the same form as in the
case $w_2=0$, except that the values of one of the $u_i$ are 
constrained (to match those of $SU(2)$, while the other is unconstrained.
Thinking of the roots above
in this
fashion can also be used to generate
table~\ref{table:so5}.

In any event, from table~\ref{table:so5}, we read off 
a one-dimensional residue integral, proportional to
\begin{eqnarray}
\lefteqn{
\frac{N}{2}  \left( \frac{2 \pi \eta(q)^3 }{ \theta_1(\tau | -z) }
\right)
\oint \frac{d u}{2\pi i}
} \nonumber \\
& & \cdot
\frac{ \theta_1( \tau | u + 1/2 ) }{ \theta_1( \tau | -z + u + 1/2) }
\frac{ \theta_1( \tau | u + \tau/2) }{ \theta_1( \tau | -z + u + \tau/2) }
\frac{ \theta_1( \tau | u - (1 + \tau)/2) }{ \theta_1( \tau | -z + u - (1 + \tau)/2)}
\nonumber \\
& & \cdot
\frac{ \theta_1( \tau | - u + 1/2 ) }{ \theta_1( \tau | -z - u + 1/2) }
\frac{ \theta_1( \tau | - u + \tau/2) }{ \theta_1( \tau | -z - u + \tau/2) }
\frac{ \theta_1( \tau | - u - (1 + \tau)/2) }{ \theta_1( \tau | -z - u - (1 + \tau)/2)},
\end{eqnarray}
where
\begin{eqnarray}
N & = &
\frac{ \theta_1( \tau | + 1/2 ) }{ \theta_1( \tau | -z + 1/2) }
\frac{ \theta_1( \tau | + \tau/2) }{ \theta_1( \tau | -z + \tau/2) }
\frac{ \theta_1( \tau | - (1 + \tau)/2) }{ \theta_1( \tau | -z - (1 + \tau)/2)},
\\
& = &
2 \frac{ \theta_1( \tau | -z ) }{
\theta_1( \tau | -2 z) },
\end{eqnarray}
where the second line follows from \cite[equ'n (A.6)]{Kim:2017zis}.

From table~\ref{table:gw}, the moduli space of flat $SO(5)$ connections
with nontrivial characteristic class is the same as the moduli space
of flat $SU(2)$ connections, i.e., $T^2 / {\mathbb Z}_2$, which is the
origin of the integral above.  We integrate over the covering space
$T^2$, and add a factor of $1/2$ (given in the expression above)
to take into account the fact that we are integrating over a double
cover of the moduli space.

Let us now evaluate this integral.
Following the Jeffrey-Kirwan residue prescription in this case,
we consider residues about three of the six poles, 
defined by
denominators with positive $u$ coefficients.  (Alternatively, we could
sum only over poles with negative $u$ coefficients, but we pick the
former convention in this paper.)
These poles are given by
\begin{equation}
u \: = \: z - 1/2, \: \: \:
z - \tau/2, \: \: \:
z + (1 + \tau)/2.
\end{equation}
The fact that the integrand is symmetric under $u \mapsto - u$
reflects the Weyl group action on the moduli space of flat
$SU(2)$ connections.
Also note that the prescription above is summing
over distinct residues which are not related by the Weyl group.

We will use the identity \cite[equ'n (B.6)]{Benini:2013nda}
\begin{equation}
\theta_1'(\tau | 0) \: = \: 2 \pi \eta(q)^3,
\end{equation}
where the derivative is taken with respect to the second variable.
As a result, 
and using the fact that \cite[equ'n (B.4)]{Benini:2013nda}
\begin{equation}
\theta_1( \tau | z + a + b \tau ) \: = \: (-)^{a+b}
\exp(- 2 \pi i b z - i \pi b^2 \tau) \theta_1(\tau | z)
\end{equation}
for $a, b \in {\mathbb Z}$,
one has \cite[equ'n (B.7)]{Benini:2013nda}
\begin{equation}
\frac{1}{2 \pi i} \int_{u = a + b \tau}
\frac{du}{ \theta_1(\tau | u) } \: = \:
(-)^{a+b} \frac{ \exp(i \pi b^2 \tau) }{
\theta_1'(\tau | 0 ) }
\: = \:
(-)^{a+b} \frac{ \exp(i \pi b^2 \tau) }{
2 \pi \eta(q)^3 },
\end{equation}
for $a, b \in {\mathbb Z}$.

From the pole at $u = z - 1/2$, we have a contribution
\begin{eqnarray}
\lefteqn{
\frac{1}{2}
\frac{ N}{\theta_1(\tau | -z)}
\theta_1(\tau | +z)
\frac{ \theta_1( \tau | z - 1/2 + \tau/2) }{
\theta_1( \tau | - 1/2 + \tau/2 ) }
\frac{ \theta_1( \tau | z - 1/2 - (1+\tau)/2 ) }{
\theta_1( \tau | - 1/2 - (1 + \tau)/2) }
} \nonumber \\
& & \cdot
\frac{ \theta_1(\tau | -z + 1/2) }{ \theta_1(\tau | -2 z) }
\frac{ \theta_1(\tau | -z + 1/2 + \tau/2) }{
\theta_1( \tau | -2z + 1/2 + \tau/2) }
\frac{ \theta_1( \tau | -z + 1/2 - (1+\tau)/2 ) }{
\theta_1( \tau | -2z + 1/2 - (1 + \tau)/2) }.
\end{eqnarray}

From the pole at $u = z - \tau/2$, we have a contribution
\begin{eqnarray}
\lefteqn{
\frac{1}{2}
\frac{ N}{\theta_1(\tau | -z)}
\theta_1(\tau | +z)
\frac{ \theta_1( \tau | z + 1/2 - \tau/2) }{
\theta_1( \tau | +1/2 - \tau/2) }
\frac{ \theta_1( \tau | z - \tau/2 - (1+\tau)/2 ) }{
\theta_1( \tau | - \tau/2 - (1+\tau)/2) }
} \nonumber \\
& & \cdot
\frac{ \theta_1( \tau | -z + \tau/2 + 1/2) }{
\theta_1( \tau | -2z + \tau/2 + 1/2) }
\frac{ \theta_1( \tau | -z + \tau) }{
\theta_1( -z + \tau) }
\frac{ \theta_1( \tau | -z - 1/2) }{
\theta_1( \tau | -2z - 1/2 ) }.
\end{eqnarray}

From the pole at $u = z + (1 + \tau)/2$, we have a contribution
\begin{eqnarray}
\lefteqn{
\frac{1}{2}
\frac{ N}{\theta_1(\tau | -z)}
\theta_1(\tau | +z)
\frac{ \theta_1( \tau | z + 1/2 + (1+\tau)/2 ) }{
\theta_1( 1/2 + (1+\tau)/2 ) }
\frac{ \theta_1( \tau | z + 1/2 + \tau) }{
\theta_1( \tau | 1/2 + \tau) }
} \nonumber \\
& & \cdot
\frac{ \theta_1( \tau | -z - \tau/2) }{
\theta_1( \tau | -2z - \tau/2) }
\frac{ \theta_1( \tau | -z - 1/2 ) }{
\theta_1( \tau | -2z - 1/2) }
\frac{ \theta_1( \tau | -z - 1 - \tau) }{
\theta_1( \tau | -2z - 1 - \tau) }.
\end{eqnarray}

One can verify (e.g. numerically) that the sum of these residues is
\begin{equation}
2
\frac{\theta_1(\tau | -z) }{ \theta_1(\tau | -2z) }
\frac{\theta_1(\tau | -3z) }{ \theta_1(\tau | -4z) }.
\end{equation}

To derive $Z(SO(5), w\neq 0)$, we still need a numerical factor,
$1/|W|$ for some $W$ as in \cite{Kologlu:2016aev}. Rather than compute
$W$ directly, for the moment,
we write
\begin{equation}
Z( SO(5), w \neq 0) \: = \: \alpha Z( SO(5), w=0)
\end{equation}
for some positive real number $\alpha$, which we will compute by using known
results for supersymmetry breaking.

Now, let us assemble these results into the elliptic genus for
$SO(5)$ with discrete theta angle $\theta \in \{0, \pi\}$.
The contribution from the sector with $w_2 = 0$ is independent of $\theta$,
and is just a factor of $1/|\Gamma|$ away from the elliptic genus of Spin$(5)$:
\begin{equation}
Z(SO(5), w = 0) \: = \:
\frac{1}{2}
\frac{\theta_1(\tau | -z) }{ \theta_1(\tau | -2z) }
\frac{\theta_1(\tau | -3z) }{ \theta_1(\tau | -4z) }.
\end{equation}

Next, we consider the contribution
from the sector with $w \neq 0$.
There is a factor of $\exp(i \theta)$ from the discrete
theta angle $\theta \in \{0, \pi\}$.  In addition, there is also a phase
$\exp(i w \cdot t)$ where, from the analysis of \cite{Gu:2020ivl},
\begin{equation}
t_a \: = \: i \pi m_a,
\end{equation}
where
\begin{equation}
\sum_a m_a \: \equiv \: 1 \mod 2.
\end{equation}
As a result, $\exp(i w \cdot t) = -1$.

Putting this together, we have the elliptic genus of a pure
supersymmetric $SO(5)$ gauge theory as a function of
discrete theta angle $\theta \in \{0, \pi\}$:
\begin{eqnarray}
Z( SO(5), \theta) & = &
Z( SO(5), w=0) \: - \: \exp(i \theta)
Z( SO(5), w \neq 0),
\\
& = &
\frac{1}{2}
\frac{\theta_1(\tau | -z) }{ \theta_1(\tau | -2z) }
\frac{\theta_1(\tau | -3z) }{ \theta_1(\tau | -4z) }
\left( 1 - \alpha \exp(i \theta) \right).
\end{eqnarray}

From \cite[section 13.2]{Gu:2018fpm}, we know that supersymmetry is broken
in pure $SO(5)$ theories with $\theta = 0$, hence we must require that
$\alpha = 1$, hence the elliptic genus of the pure $SO(5)$ theory with
discrete theta angle $\theta$ is
\begin{equation}
Z( SO(5), \theta)
\: = \:
\frac{1}{2}
\frac{\theta_1(\tau | -z) }{ \theta_1(\tau | -2z) }
\frac{\theta_1(\tau | -3z) }{ \theta_1(\tau | -4z) }
\left( 1 - \exp(i \theta) \right).
\end{equation}

As a consistency check, note that $\alpha$ is a positive real
number, as expected -- phase factors have already been accounted for.
As another consistency check, 
note that for $\theta = \pi$, the elliptic genus of
the pure $SO(5)$ gauge theory matches that of the Spin$(5)$ theory,
in agreement with expectations from \cite[section 13.2]{Gu:2018fpm}.

As another consistency check, note that this implies that the elliptic
genus of the pure Spin$(5)$ theory is the sum of the elliptic genera
of the pure $SO(5)$ theories with either value of $\theta$:
\begin{equation}
{\rm Spin}(5) \: = \: SO(5)_{\theta = 0 } \: + \:
SO(5)_{\theta = \pi},
\end{equation}
which is consistent with decomposition of two-dimensional theories
with a $B {\mathbb Z}_2$ symmetry 
\cite{Hellerman:2006zs,Sharpe:2014tca,Sharpe:2019ddn}.

\section{Pure $Sp(6)/{\mathbb Z}_2$ gauge theories}
\label{sect:sp6}

We now turn to pure $Sp(6)/{\mathbb Z}_2$ gauge theories
(in conventions in which $Sp(2) = SU(2)$).  
Since $Sp(2) = SU(2)$ and $Sp(4) = {\rm Spin}(5)$, the first
interesting case amongst $Sp(2k)/{\mathbb Z}_2$ is $Sp(6)/{\mathbb Z}_2$.

As before, for bundles of vanishing characteristic class,
from equation~(\ref{eq:factor}),
the elliptic genus matches that of the pure $Sp(6)$ gauge theory,
up to the factor $1/|\Gamma|$:
\begin{equation}
Z( Sp(6)/{\mathbb Z}_2, w=0) \: = \: 
\frac{1}{2} Z( Sp(6) ) \: = \:
\frac{1}{2}
\frac{ \theta_1( \tau | -z) }{ \theta_1( \tau | -2z) }
\frac{ \theta_1( \tau | -3z) }{ \theta_1( \tau | -4z) }
\frac{ \theta_1( \tau | -5z) }{ \theta_1( \tau | -6z) },
\end{equation}
as discussed in section~\ref{sect:strategy}.

To describe a nontrivial
bundle, we give two anticommuting holonomies in $Sp(2k)$,
which following \cite[section 4.1]{Witten:1997bs}, 
\cite[equ'n (8)]{Keurentjes:2000bs} we can take
to be 
\begin{eqnarray}
p & = & {\rm diag}\left( \lambda_1, - \lambda_1, i, -i, 
- \lambda_1^{-1}, \lambda_1^{-1} \right),
\\
q & = & {\rm diag}\left( 
\left[ \begin{array}{cc}
0 & - \lambda_2 \\
-\lambda_2 & 0 \end{array} \right],
\left[ \begin{array}{cc}
0 & -i \\
-i & 0 \end{array} \right],
\left[ \begin{array}{cc}
0 & -\lambda_2^{-1} \\
- \lambda_2^{-1} & 0 
\end{array} \right] \right),
\end{eqnarray}
and where we take the symplectic form to be
\begin{equation}
\label{eq:symplecticform}
\Omega \: = \: \left[ \begin{array}{cccccc}
0 & 0 & 0 & 0 & 0 & -1 \\
0 & 0 & 0 & 0 & -1 & 0 \\
0 & 0 & 0 & -1 & 0 & 0 \\
0 & 0 & 1 & 0 & 0 & 0 \\
0 & 1 & 0 & 0 & 0 & 0 \\
1 & 0 & 0 & 0 & 0 & 0
\end{array} \right],
\end{equation}
so that
\begin{equation}
p^T \Omega p \: = \: \Omega,
\: \: \:
q^T \Omega q \: = \: \Omega.
\end{equation}

Following the procedure
of section~\ref{sect:strategy}, we diagonalize a basis of the Lie algebra
\footnote{
In case the reader finds it helpful, the Lie algebra with the symplectic form given in Equation~\eqref{eq:symplecticform} is described in detail in 
\cite[Chapter 30]{Bump:2013}.
}
with respect to the diagonal action of $p$, $q$ above.
The eigenvalues $\omega_{p,q}^{\alpha}$ of the
adjoint action 
are given in table~\ref{table:sp3}.

\begin{table}
\begin{center}
\begin{tabular}{cc|c} 
$\omega_p$ & $\omega_q$ & $\theta$ argument\\ \hline
$\lambda_1^{-2}$ & $- \lambda_2^{-2}$ & $\tau/2 - 2 u$ \\
$\lambda_1^{-2}$ & $\lambda_2^{-2}$ & $-2u$ \\
$-\lambda_1^{-2}$ & $\lambda_2^{-2}$ & $1/2 - 2 u$ \\
$-i \lambda_1^{-1}$ & $-i \lambda_2^{-1}$ & $3/4 + 3\tau/4 - u$ \\
$i \lambda_1^{-1}$ & $-i \lambda_2^{-1}$ & $1/4 + 3 \tau/4 - u$ \\
$-i \lambda_1^{-1}$ & $i \lambda_2^{-1}$ & $3/4 + \tau/4 - u$ \\
$i \lambda_1^{-1}$ & $i \lambda_2^{-1}$ & $1/4 + \tau/4 - u$ \\
$-1$ & $-1$ & $-(1 + \tau)/2$ \\
$-1$ & $-1$ & $-(1 + \tau)/2$ \\
$1$ & $-1$ & $\tau/2$ \\
$1$ & $-1$ & $\tau/2$ \\
$-1$ & $1$ & $1/2$ \\
$-1$ & $1$ & $1/2$ \\
$1$ & $1$ & $0$ \\
$-i \lambda_1$ & $-i \lambda_2$ & $3/4 + 3 \tau/4 + u$ \\
$i \lambda_1$ & $-i \lambda_2$ & $1/4 + 3 \tau/4 + u$ \\
$-i \lambda_1$ & $i \lambda_2$ & $3/4 + \tau/4 + u$ \\
$i \lambda_1$ & $i \lambda_2$ & $1/4 + \tau/4 + u$ \\
$\lambda_1^2$ & $-\lambda_2^2$ & $\tau/2 + 2 u$ \\
$\lambda_1^2$ & $\lambda_2^2$ & $2u$ \\
$-\lambda_1^2$ & $\lambda_2^2$ & $1/2 + 2u$
\end{tabular}
\end{center}
\caption{Table of eigenvalues of the adjoint action of the holonomy
matrices. \label{table:sp3} }
\end{table}

In table~\ref{table:sp3},
\begin{equation}
u \: = \: \frac{\ln \lambda_1}{2 \pi i} \: + \: \tau \frac{ \ln \lambda_2}{
2 \pi i},
\end{equation}
and the $\theta$ coefficient is
\begin{equation}
\frac{\ln \omega_p^{\alpha}}{2\pi i} \: + \: \tau 
\frac{\ln \omega_q^{\alpha}}{2\pi i}.
\end{equation}
As a simple consistency check, note that
the number of entries, 21, is the same as the dimension of $Sp(6)$.

The fact that there is only one entry in table~\ref{table:sp3} with
$p$, $q$ eigenvalues $(1,1)$ means that the elliptic genus will be computed
by a one-dimensional residue integral.  From table~\ref{table:gw},
we see that the moduli space of flat $Sp(6)/{\mathbb Z}_2$ connections
with nontrivial characteristic class is the same as the moduli space
of $SU(2)$ connections -- indeed, one-dimensional.  The moduli space
is $T^2/{\mathbb Z}_2$, but we will integrate over the $T^2$ cover,
quotienting by a factor of $2$ to reflect that fact.

Putting this together, the elliptic genus of a pure $Sp(6)/{\mathbb Z}_2$ gauge
theory with bundles of nontrivial characteristic class is proportional to
\begin{eqnarray} \label{eq:sp6z2int}
\lefteqn{
\frac{N}{2} \left( \frac{2 \pi \eta(q)^3}{\theta_1(\tau | -z)} \right)
\oint \frac{du}{2\pi i}
}  \\
& & \cdot
\frac{ \theta_1( \tau | 3/4 + 3 \tau/4 + u )}{
\theta_1( -z +  3/4 + 3 \tau/4 + u ) }
\frac{ \theta_1(\tau | 1/4 + 3 \tau/4 + u) }{
\theta_1(\tau | -z +  1/4 + 3 \tau/4 + u) }
\frac{ \theta_1( \tau | 3/4 + \tau/4 + u) }{
\theta_1( \tau | -z +  3/4 + \tau/4 + u) }
\nonumber \\
& & \cdot
\frac{ \theta_1( \tau | 1/4 + \tau/4 + u) }{
\theta_1(\tau | -z +  1/4 + \tau/4 + u) }
\frac{ \theta_1( \tau | \tau/2 + 2u) }{
\theta_1( \tau | -z +  \tau/2 + 2u) }
\frac{ \theta_1( \tau | 2u ) }{
\theta_1( \tau | -z + 2u) }
\frac{ \theta_1( \tau | 1/2 + 2u) }{
\theta_1( -z + 1/2 + 2u) }
\nonumber \\   
& & \cdot
\frac{ \theta_1( \tau | 3/4 + 3 \tau/4 - u )}{
\theta_1( -z +  3/4 + 3 \tau/4 - u ) }
\frac{ \theta_1(\tau | 1/4 + 3 \tau/4 - u) }{
\theta_1(\tau | -z +  1/4 + 3 \tau/4 - u) }
\frac{ \theta_1( \tau | 3/4 + \tau/4 - u) }{
\theta_1( \tau | -z +  3/4 + \tau/4 - u) }
\nonumber \\
& & \cdot
\frac{ \theta_1( \tau | 1/4 + \tau/4 - u) }{
\theta_1(\tau | -z +  1/4 + \tau/4 - u) }
\frac{ \theta_1( \tau | \tau/2 - 2u) }{
\theta_1( \tau | -z +  \tau/2 - 2u) }
\frac{ \theta_1( \tau | -2u ) }{
\theta_1( \tau | -z - 2u) }
\frac{ \theta_1( \tau | 1/2 - 2u) }{
\theta_1( -z + 1/2 - 2u) },
\nonumber
\end{eqnarray}
where
\begin{eqnarray}
N & = & \left[
\frac{ \theta_1( \tau | - (1 + \tau)/2 )}{ 
\theta_1( \tau | -z - (1 + \tau)/2 )}
\frac{ \theta_1( \tau | \tau/2 ) }{
\theta_1( \tau | -z + \tau/2 ) }
\frac{ \theta_1( \tau | 1/2) }{
\theta_1( \tau | -z + 1/2) }
\right]^2,
\\
& = & \left[
2 
\frac{ \theta_1( \tau | -z) }{ \theta_1(\tau | -2z) }
\right]^2,
\end{eqnarray}
using \cite[appendix A]{Kim:2017zis}.
The overall factor of $1/2$ is due to the fact that we are integrating
over the double-cover $T^2$ of the moduli space of flat connections.

The reader will note that the expression above is symmetric under
$u \leftrightarrow -u$.  This reflects the Weyl group action on the
moduli space of flat $SU(2)$ connections, whose double-cover we are
integrating over in the expression above.

Following the Jeffrey-Kirwan residue prescription, we will take poles of denominators
in which $u$ appears with a positive coefficient.
(Alternatively, we could equivalently 
take poles in which $u$ appears with negative
coefficient, but we will use the positive coefficient prescription in this
paper.)  In passing, note that none of these poles are related
by the Weyl group action to one another.

Four of the poles are at
\begin{equation}
u \: = \: z - 3/4 - 3\tau/4,
\: \: \:
z - 1/4 - 3 \tau/4,
\: \: \:
z - 3/4 - \tau/4,
\: \: \:
z - 1/4 - \tau/4.
\end{equation}
To find all of the remaining poles, one must take into account the
periodicities of the theta function.
Taking those into account, we find four poles at
\begin{equation}
2u \: = \: z - \tau/2 + \{0, 1, \tau, 1+\tau \},
\mbox{  or  }
u \: = \: z/2 - \tau/4 + \{0, 1/2, \tau/2, (1+\tau)/2)\},
\end{equation}
another four poles at
\begin{equation}
u \: = \: z/2 + \{0, 1/2, \tau/2, (1+\tau)/2)\},
\end{equation}
and another four at
\begin{equation}
u \: = \: z/2 - 1/4 + \{0, 1/2, \tau/2, (1+\tau)/2)\},
\end{equation}
for a total of 16 residues that must be summed over.

We illustrate a few examples of these residues here, to illustrate
the complexity of the computation.
The residue at $u = z - 3/4 - 3\tau/4$ is given by
\begin{eqnarray}
\lefteqn{
\frac{1}{2}
\frac{N}{\theta_1(\tau | -z)}
 \theta_1( \tau | +z)
\frac{ \theta_1( \tau | z - 1/2) }{ \theta_1( \tau | -1/2) }
\frac{ \theta_1( \tau | z - \tau/2) }{ \theta_1( \tau | - \tau/2) }
\frac{ \theta_1( \tau | z - (1 + \tau)/2 ) }{
\theta_1( \tau | - (1+\tau)/2 ) }
} \nonumber \\
& & \cdot
\frac{ \theta_1( \tau | 2z - 3/2 - \tau) }{
\theta_1( \tau | z - 3/2 - \tau) }
\frac{ \theta_1( \tau | 2z - 3/2 - 3\tau/2) }{
\theta_1( \tau | z - 3/2 - 3\tau/2) }
\frac{ \theta_1( \tau | 2z - 1 - 3\tau/4) }{
\theta_1( \tau | z -   3\tau/4) }
\nonumber \\
& & \cdot
\frac{ \theta_1( \tau | -2z + 3/2 + 2 \tau) }{
\theta_1( \tau | -3z + 3/2 + 2 \tau) }
\frac{ \theta_1( \tau | -2z + 3/2 + 3 \tau/2) }{
\theta_1(\tau | -3z + 3/2 + 3\tau/2) }
\frac{ \theta_1( \tau | -2z + 2 + 3\tau/2) }{
\theta_1( \tau | -3z + 2 + 3 \tau/2) }
\nonumber \\
& & \cdot
\frac{ \theta_1( \tau | -z + 3/2 + 3\tau/2) }{
\theta_1( \tau | -2z + 3/2 + 3\tau/2) }
\frac{ \theta_1( \tau | -z + 1 + 3\tau/2) }{
\theta_1( \tau | -2z + 1 + 3 \tau/2) }
\frac{ \theta_1( \tau | -z + 3/2 + \tau) }{
\theta_1( \tau | -2z + 3/2 + \tau) }
\nonumber \\
& & \cdot
\frac{ \theta_1( \tau | -z + 1 + \tau) }{
\theta_1( \tau | -2z + 1 + \tau) }.
\end{eqnarray}
Similarly, the residue at $u = z/2 - \tau/4$ is given by
\begin{eqnarray}
\lefteqn{
\frac{1}{4} \frac{N}{\theta_1(\tau | -z)}
\frac{ \theta_1( \tau | z/2 + 3/4 + \tau/2) }{
\theta_1( \tau | - z/2 + 3/4 + \tau/2) }
\frac{ \theta_1( \tau | z/2 + 1/4 + \tau/2) }{
\theta_1( \tau | - z/2 + 1/4 + \tau/2) }
\frac{ \theta_1( \tau | z/2 + 3/4) }{
\theta_1( \tau | -z/2 + 3/4) }
} \nonumber \\
& & \cdot
\frac{ \theta_1( \tau | z/2 + 1/4) }{
\theta_1( \tau | - z/2 + 1/4) }
\theta_1(+z)
\frac{\theta_1( \tau | z/2 - \tau/2) }{
\theta_1( \tau | - \tau/2) }
\frac{ \theta_1( \tau | z + 1/2 - \tau/2) }{
\theta_1(\tau | 1/2 - \tau/2) }
\nonumber \\
& & \cdot
\frac{\theta_1( \tau | -z + \tau) }{ 
\theta_1( \tau | -2z + \tau) }
\frac{ \theta_1( \tau | -z + \tau/2) }{
\theta_1( \tau | -2z + \tau/2) }
\frac{ \theta_1( \tau | -z + 1/2 + \tau/2) }{
\theta_1( \tau | -2z + 1/2 + \tau/2) }
\frac{ \theta_1( \tau | - z/2 + 3/4 + \tau) }{
\theta_1( \tau | - 3 z/2 + 3/4 + \tau) }
\nonumber \\
& & \cdot
\frac{\theta_1(\tau | - z/2 + 1/4 + \tau) }{
\theta_1( \tau | - 3 z / 2 + 1/4 + \tau) }
\frac{ \theta_1( \tau | - z/2 + 3/4 + \tau/2) }{
\theta_1( \tau | - 3 z/2 + 3/4 + \tau/2) }
\frac{ \theta_1( \tau | - z/2 + 1/4 + \tau/2) }{
\theta_1( \tau | - 3 z/2 + 1/4 + \tau/2) }.
\end{eqnarray}
A leading factor of $1/2$ in the second residue
is due to the fact that the pole arises
from a theta function denominator that depends upon $2u$ not $u$.
An overall factor of $1/2$ in both residues is due to the fact that
we are integrating over $T^2$ and not $T^2/{\mathbb Z}_2$.
For reasons of brevity, we do not list the other fourteen residues here,
though they are straightforward to compute.

One can verify numerically that the sum of the residues above,
the integral~(\ref{eq:sp6z2int}) 
equals
\begin{equation}
8 
\frac{ \theta_1( \tau | -z) }{ \theta_1( \tau | -2z) }
\frac{ \theta_1( \tau | -3z ) }{ \theta_1( \tau | -4z) }
\frac{ \theta_1( \tau | -5z) }{ \theta_1( \tau | -6z ) }.
\end{equation}

The product of theta functions above should be proportional to
the elliptic genus of the pure $Sp(6)/{\mathbb Z}_2$ theory with
nonzero characteristic class.  The proportionality factor
should be a real number of the form $1/|W|$ for $W$ a finite subgroup
of the gauge group that preserves the holonomies.  
For the moment, we will write
\begin{equation}
Z( Sp(6)/{\mathbb Z}_2, w \neq 0) \: = \:
\alpha Z( Sp(6)/{\mathbb Z}_2, w = 0),
\end{equation}
for some positive real number $\alpha$.
We will compute this
factor indirectly, using known results for supersymmetry breaking
for various discrete theta angles.

Finally, we need to weight the $w \neq 0$ contribution with relevant
phases.  There is a factor $\exp(i \theta)$ arising from the discrete
theta angle $\theta \in \{0, \pi\}$.  In addition, there is potentially
a factor of $\exp(i w \cdot t)$.  From 
\cite[section 5]{Gu:2020ivl},
\begin{equation}
t_a \: = \: \pi i m_a
\end{equation}
where
\begin{equation}
\sum_a m_a  \: \equiv \: 0 \mod 2
\end{equation}
(for $Sp(6)/{\mathbb Z}_2$),
so without loss of generality we can take all $m_a = 0$, hence
$\exp(i w \cdot t) = +1$.

Now, putting this together, combining the result for the elliptic genus in
the sector with
$w=0$ with the result above, determined up to a proportionality factor,
for $w \neq 0$, we have
that the elliptic genus of a pure $Sp(6)/{\mathbb Z}_2$ gauge
theory with discrete theta angle $\theta \in \{0, \pi\}$ is given by
\begin{eqnarray}
Z( Sp(6)/{\mathbb Z}_2, \theta) & = &
Z( Sp(6)/{\mathbb Z}_2, w=0) \: + \: \alpha \exp(i \theta) 
Z( Sp(6)/{\mathbb Z}_2, w \neq 0),
\\
& = &
\frac{1}{2} \frac{ \theta_1( \tau | -z) }{ \theta_1( \tau | -2z) }
\frac{ \theta_1( \tau | -3z) }{ \theta_1( \tau | -4z) }
\frac{ \theta_1( \tau | -5z) }{ \theta_1( \tau | -6z) }
\left( 1 \: + \: \alpha \exp(i \theta) \right).
\end{eqnarray}

It was argued in \cite[section 5]{Gu:2020ivl}
that a pure $Sp(6)/{\mathbb Z}_2$
gauge theory has supersymmetric vacua only if the discrete theta angle
$\theta = 0$, hence for $\theta = \pi$, supersymmetry is broken, and the
elliptic genus should vanish.  Imposing this as a constraint, we find that
$\alpha = +1$, hence the elliptic genus of a pure $Sp(6)/{\mathbb Z}_2$
gauge theory as a function of discrete theta angle $\theta$ is
\begin{equation}
Z( Sp(6)/{\mathbb Z}_2, \theta) \: = \:
\frac{1}{2} \frac{ \theta_1( \tau | -z) }{ \theta_1( \tau | -2z) }
\frac{ \theta_1( \tau | -3z) }{ \theta_1( \tau | -4z) }
\frac{ \theta_1( \tau | -5z) }{ \theta_1( \tau | -6z) }
\left( 1 \: + \: \exp(i \theta) \right).
\end{equation}

As a consistency check, note that $\alpha$ is real and positive, as expected --
phase factors have already been accounted for.
As another consistency check, note that for $\theta = 0$, the
elliptic genus of the pure $Sp(6)/{\mathbb Z}_2$ gauge theory matches
that of the pure $Sp(6)$ gauge theory, in agreement with expectations
from \cite[section 5]{Gu:2020ivl}.

As a further consistency check, it is straightforward to see that this result
is consistent with decomposition \cite{Hellerman:2006zs,Sharpe:2014tca,Sharpe:2019ddn}:
\begin{equation}
\sum_{\theta = 0, \pi}
\frac{1}{2} \frac{ \theta_1( \tau | -z) }{ \theta_1( \tau | -2z) }
\frac{ \theta_1( \tau | -3z) }{ \theta_1( \tau | -4z) }
\frac{ \theta_1( \tau | -5z) }{ \theta_1( \tau | -6z) }
\left( 1 \: + \: \exp(i \theta) \right)
\: = \: Z( Sp(6) ),
\end{equation}
consistent with the expectation
\begin{equation}
Sp(6) \: = \: \left( Sp(6)/{\mathbb Z}_2 \right)_{\theta = 0}
\: + \:
\left( Sp(6)/{\mathbb Z}_2 \right)_{\theta = \pi}
\end{equation}
(expressed schematically).

\section{Predictions for general cases}
\label{sect:predict}

So far, we have performed direct computations to compute elliptic
genera of pure gauge theories with semisimple, non-simply-connected
gauge groups in some low rank cases.  Next, we are going to make a proposal
for all cases, utilizing (a) our knowledge of the contribution from
$w=0$, (b) supersymmetry breaking for most discrete theta angles, and
(c) decomposition.
These three constraints form sufficiently many algebraic
equations to enable us to solve algebraically for the elliptic genera.

We illustrate
the method using the pure $SU(4)/{\mathbb Z}_4$
gauge theory as an example.  First, we know that
\begin{equation}
Z( SU(4)/{\mathbb Z}_4, w=0) \: = \: \frac{1}{4}
Z( SU(4) ).
\end{equation}
Given the results for low-rank cases, let us assume that
\begin{equation}
Z( SU(4)/{\mathbb Z}_4, w\neq 0) \: \propto \:
Z( SU(4)/{\mathbb Z}_4, w=0),
\end{equation}
so we can write
\begin{equation}
Z( SU(4)/{\mathbb Z}_4, \theta) \: = \:
 \frac{1}{4} Z( SU(4) ) \left(
1 \: + \: \alpha_1 \exp(i \theta) \: + \:
\alpha_2 \exp(2 i \theta) \: + \: \alpha_3 \exp(3 i \theta) \right),
\end{equation}
for $\theta \in \{0, \pi/2, \pi, 3\pi/2\}$.
From table~\ref{table:dta},we see that supersymmetry is broken unless 
$\theta = \pi$, which gives the constraints
\begin{eqnarray}
1 \: + \: \alpha_1 \: + \: \alpha_2 \: + \: \alpha_3 & = & 0,
\\
1 \: + \: i \alpha_1 \: - \: \alpha_2 \: - \: i \alpha_3 & = & 0,
\\
1 \: - \: i \alpha_1 \: - \: \alpha_2 \: + \: i \alpha_3 & = & 0,
\end{eqnarray}
for $\theta = 0, \pi/2, 3\pi/2$, respectively,
and from decomposition, since the elliptic genera vanish for
$\theta \neq \pi$, the elliptic genus at $\theta = \pi$ must match
that of $SU(4)$, hence
\begin{equation}
1 \: - \: \alpha_1 \: + \: \alpha_2 \: - \: \alpha_3 \: = \: 4.
\end{equation}
These are four linear algebraic equations in three unknowns,
which happen to admit a unique solution:
\begin{equation}
\alpha_1 \: = \: \alpha_3 \: = \: -1, \: \: \:
\alpha_2 \: = \: +1.
\end{equation}
Putting this together, we have that
\begin{equation}
Z(SU(4)/{\mathbb Z}_4, \theta) \: = \:
\frac{1}{4} \frac{ \theta_1(\tau | -z) }{ \theta_1( \tau | -4z) }
\left( 1 \: - \: \exp(i \theta) \: + \: \exp(2 i \theta) \: - \:
\exp(3 i \theta) \right).
\end{equation}
We have used our knowledge of supersymmetry breaking and decomposition,
and only assumed that the contributions from sectors of various
characteristic classes are proportional to one another.  One can check
that the resulting phase factors, derived algebraically, are consistent
with those described in section~\ref{sect:strategy}.

Proceeding in this fashion, using our knowledge of supersymmetry breaking
and decomposition, elliptic genera are straightforward to predict for all
other cases.  We summarize the results below.

First, for $SU(k)/{\mathbb Z}_k$, for $k$ odd, supersymmetry is unbroken
only for $\theta = 0$ (from table~\ref{table:dta}), 
and we predict the elliptic genus
\begin{equation}
Z( SU(k)/{\mathbb Z}_k, \theta) \: = \: 
\frac{1}{k} \frac{ \theta_1( \tau | -z) }{ \theta_1( \tau | -k z) }
\sum_{m=0}^{k-1} \exp( i m \theta),
\end{equation}
for $\theta \in \{0, 2\pi/k, 4 \pi/k, \cdots, 2 (k-1) \pi/k \}$.
For $k$ even, supersymmetry is unbroken only for $\theta = \pi$
and we predict the elliptic genus
\begin{equation}
Z( SU(k)/{\mathbb Z}_k, \theta) \: = \: 
\frac{1}{k} \frac{ \theta_1( \tau | -z) }{ \theta_1( \tau | -k z) }
\sum_{m=0}^{k-1} (-)^m \exp( i m \theta).
\end{equation}

Proceeding similarly,
for Spin$(2k+1)/{\mathbb Z}_2$, we predict the elliptic genus
\begin{equation}
Z( {\rm Spin}(2k+1)/{\mathbb Z}_2, \theta) \: = \:
\frac{1}{2} Z( {\rm Spin}(2k+1) ) \left( 1 - \exp(i \theta) \right),
\end{equation}
where $Z( {\rm Spin}(2k+1) )$ denotes the elliptic genus of the
pure Spin$(2k+1)$ gauge theory, as given in
section~\ref{sect:rev}, and for $\theta \in \{0, \pi\}$.

For Spin$(4k)/{\mathbb Z}_2 \times {\mathbb Z}_2$, 
we predict the
elliptic genus
\begin{equation}
Z( {\rm Spin}(4k)/{\mathbb Z}_2 \times {\mathbb Z}_2, \theta_1, \theta_2)
\: = \:
\frac{1}{4} Z( {\rm Spin}(4k) ) \left( 1 + (-)^k \exp(i \theta_1) \right)
\left( 1 + (-)^k \exp( i \theta_2) \right),
\end{equation}
for $\theta_{1,2} \in \{0, \pi\}$.

For Spin$(4k+2)/{\mathbb Z}_4$, we predict the elliptic genus
\begin{equation}
Z( {\rm Spin}(4k+2)/{\mathbb Z}_4, \theta)
\: = \: 
\frac{1}{4} Z( {\rm Spin}(4k+2) )
\sum_{m=0}^3 (-)^{km} \exp(i m \theta),
\end{equation}
for $\theta \in \{0, \pi\}$.

For $Sp(2k)/{\mathbb Z}_2$, we predict the elliptic genus
\begin{equation}
Z( Sp(2k)/{\mathbb Z}_2, \theta) \: = \:
\frac{1}{2} Z( Sp(2k) ) \left( 1 + (-)^m \exp(i \theta) \right),
\end{equation}
for $\theta \in \{0, \pi\}$,
where 
\begin{equation}
m \: = \: \left\{ \begin{array}{cl}
k/2 & k \: {\rm even}, \\
(k+1)/2 & k \: {\rm odd}.
\end{array} \right.
\end{equation}

For $E_6/{\mathbb Z}_3$, we predict the elliptic genus
\begin{equation}
Z( E_6/{\mathbb Z}_3, \theta) \: = \:
\frac{1}{3} Z( E_6 ) \left( 1 + \exp(i \theta) + \exp(2 i \theta) \right),
\end{equation}
for $\theta \in \{0, 2 \pi/3, 4 \pi/3 \}$.

For $E_7/{\mathbb Z}_2$, we predict the elliptic genus
\begin{equation}
Z( E_7/ {\mathbb Z}_2, \theta) \: = \:
\frac{1}{2} Z( E_7 ) \left( 1 - \exp(i \theta) \right),
\end{equation}
for $\theta \in \{0, \pi\}$.

As a consistency check, note that the elliptic genus of
$SU(2)/{\mathbb Z}_2$ matches that of Spin$(3)/{\mathbb Z}_2$,
the elliptic genus of $SU(4)/{\mathbb Z}_4$ matches that of
Spin$(6)/{\mathbb Z}_4$,  and the elliptic genus of
$Sp(4)/{\mathbb Z}_2$ matches that of Spin$(5)/{\mathbb Z}_2$,
as expected since the Lie groups are the same.

In each case, the elliptic genus vanishes for discrete theta angles $\theta$
for which supersymmetry is broken in the IR (from table~\ref{table:dta}),
and decomposition \cite{Hellerman:2006zs,Sharpe:2014tca,Sharpe:2019ddn} 
is obeyed:
\begin{equation}
Z( G ) \: = \: \sum_{\theta} Z( G/\Gamma, \theta).
\end{equation}

\section{Conclusions}

In this paper we have described
a systematic method to compute elliptic genera
of pure two-dimensional (2,2) supersymmetric $G/\Gamma$ gauge theories
with various discrete theta angles.
Our results agree with previous
computations of elliptic genera of pure $SO(3)$ gauge theories,
and we also derived the elliptic genera of pure  $SU(3)/{\mathbb Z}_3$,
$SO(4)$, Spin$(4)/{\mathbb Z}_2 \times {\mathbb Z}_2$, $SO(5)$
and $Sp(6)/{\mathbb Z}_2$ gauge theories.  
In each case, 
the results are consistent with predictions of supersymmetry
breaking for certain discrete theta angles in 
\cite{Gu:2018fpm,Chen:2018wep,Gu:2020ivl}, and 
the resulting elliptic genera are also consistent with expectations from
decomposition \cite{Hellerman:2006zs,Sharpe:2014tca,Sharpe:2019ddn} 
of two-dimensional gauge theories with
finite global one-form symmetries.  Finally, we applied these two
criteria to make predictions for elliptic genera of higher-rank cases.

Pure two-dimensional (2,2) supersymmetric gauge theories have also
been extensively studied by lattice simulations \cite{Cohen:2003xe, Kanamori:2007yx, Kanamori:2010gw, Hanada:2017gqc, Catterall:2017xox}.
Our results also provide new analytic results that can be used to test and callibrate future lattice studies
of pure two-dimensional supersymmetric gauge theories.  They also suggest new avenues for research such as varying the global structure of the gauge group and including discrete theta angles.

Gauge theories correspond to
sigma models on stacks \cite{Pantev:2005rh,Pantev:2005zs,Pantev:2005wj},
and the elliptic genera we have computed in this paper
should correspond to elliptic genera of the classifying stacks $BG$
\cite{gro,bet}.

The sensitivity of the elliptic genus to the global structure of the gauge group makes it a powerful tool to investigate of two-dimensional dualities.
The elliptic genus has already been used to test several of Hori's 
proposed dualities 
\cite{Hori:2011pd} in \cite{Kim:2017zis,Avraham:2019uiz,Bergman:2018vqe}.
Looking forward, we expect the elliptic genus of $G/\Gamma$ gauge theories will be useful to establish new dualities and will help with exploring the dynamics of two-dimensional supersymmetric gauge theories
\footnote{That said, elliptic genera should be applied with care.  For example,
we have seen earlier in this paper that the $SU(2)$ elliptic genus matches
that of $SO(3)_-$.  However, these two theories are not dual to one another.
Instead, the $SU(2)$ theory is a sum of the two $SO(3)$ theories, with
each value of the discrete theta angle.  Because supersymmetry is broken
in the $SO(3)_{+}$ theory, the elliptic genus only receives contributions
from the $SO(3)_{-}$ theory.  We see that relying solely upon the equality of elliptic genera can be misleading in
trying to find dualities.}.

\section{Acknowledgements}

We would like to thank D.~Berwick-Evans,
C.~Closset, M.~Hanada, T.~Johnson-Freyd, K.~Hori, E.~Poppitz, S.~Razamat, 
Y.~Tachikawa, A.~Tripathy, and Piljin~Yi for useful discussions.  We would especially like to thank Y.~Tachikawa for his careful reading of the manuscript.
R.E. would like to thank Kavli IPMU for hospitality while this work was being completed and the World Premier International Research Center Initiative (WPI), MEXT, Japan.
R.E. is supported in part by KIAS Individual Grant PG075901.
E.S. was partially supported by NSF grant PHY-1720321.

\end{document}